\documentclass[superscriptaddress,aps,prx,reprint,twocolumn,amsmath,amssymb,showpacs,]{revtex4-2}

\usepackage{bm}
\usepackage{bm,dcolumn,amsmath,graphicx,natbib,url,textcase}
\usepackage{epsfig}
\usepackage{epstopdf}
\usepackage{euscript}
\usepackage{amsfonts}
\usepackage{float}
\usepackage{multirow}
\usepackage{braket}
\usepackage{eufrak}
\DeclareMathAlphabet{\mathpzc}{OT1}{pzc}{m}{it}
\usepackage{threeparttable}
\usepackage{subfigure}
\usepackage{graphics}     
\usepackage{graphicx}     
\usepackage{longtable}     
\usepackage{url}          
\usepackage{bm}           
\usepackage{natbib}
\usepackage{textpos}
\usepackage{amsfonts,amsmath,amssymb,mathrsfs,bm,amsthm}
\usepackage{float}
\usepackage{booktabs}
\usepackage{array}
\usepackage{tabu}
\usepackage{dcolumn}
\usepackage{rotating}
\usepackage{ulem}
\usepackage{soul}
\usepackage{natbib}
\usepackage{upgreek}
\usepackage{textcomp}
\usepackage{xcolor}
\DeclareMathAlphabet{\mathpzc}{OT1}{pzc}{m}{it}

\newcommand{\be}{\begin{equation}}
\newcommand{\ee}{\end{equation}}

\newcommand{\RN}[1]{%
  \textup{\uppercase\expandafter{\romannumeral#1}}%
}

\usepackage{nicefrac}
\usepackage{amsthm}
\usepackage{color}
\usepackage{cancel}
\usepackage{appendix}
\usepackage{makecell}
\usepackage{xcolor}
\usepackage[colorlinks=true,linkcolor=black,citecolor=blue,urlcolor=blue,bookmarksopen]{hyperref}

\newcommand{\appropto}{\mathrel{\vcenter{
  \offinterlineskip\halign{\hfil$##$\cr
    \propto\cr\noalign{\kern2pt}\sim\cr\noalign{\kern-2pt}}}}}

\begin{document}
\title{\Large{Searches for exotic spin-dependent interactions with spin sensors}}

\date{\today}

\author{Min Jiang}
\email[]{These authors contributed equally to this work}
\affiliation{
CAS Key Laboratory of Microscale Magnetic Resonance and School of Physical Sciences, University of Science and Technology of China, Hefei, Anhui 230026, China}
\affiliation{
CAS Center for Excellence in Quantum Information and Quantum Physics, University of Science and Technology of China, Hefei, Anhui 230026, China}

\author{Haowen Su}
\email[]{These authors contributed equally to this work}
\affiliation{
CAS Key Laboratory of Microscale Magnetic Resonance and School of Physical Sciences, University of Science and Technology of China, Hefei, Anhui 230026, China}
\affiliation{
CAS Center for Excellence in Quantum Information and Quantum Physics, University of Science and Technology of China, Hefei, Anhui 230026, China}

\author{Yifan Chen}
\email[]{These authors contributed equally to this work}
\affiliation{
\mbox{Niels Bohr International Academy, Niels Bohr Institute, Blegdamsvej 17, Copenhagen 2100, Denmark}}

\author{Man Jiao}
\affiliation{Institute of Quantum Sensing and School of Physics, Zhejiang University, Hangzhou 310027, China}
\affiliation{Institute for Advanced Study in Physics, Zhejiang University, Hangzhou 310027, China}

\author{Ying Huang}
\affiliation{
CAS Key Laboratory of Microscale Magnetic Resonance and School of Physical Sciences, University of Science and Technology of China, Hefei, Anhui 230026, China}
\affiliation{
CAS Center for Excellence in Quantum Information and Quantum Physics, University of Science and Technology of China, Hefei, Anhui 230026, China}

\author{Yuanhong Wang}
\affiliation{
CAS Key Laboratory of Microscale Magnetic Resonance and School of Physical Sciences, University of Science and Technology of China, Hefei, Anhui 230026, China}
\affiliation{
CAS Center for Excellence in Quantum Information and Quantum Physics, University of Science and Technology of China, Hefei, Anhui 230026, China}

\author{Xing Rong}
\affiliation{
CAS Key Laboratory of Microscale Magnetic Resonance and School of Physical Sciences, University of Science and Technology of China, Hefei, Anhui 230026, China}
\affiliation{
CAS Center for Excellence in Quantum Information and Quantum Physics, University of Science and Technology of China, Hefei, Anhui 230026, China}

\author{Xinhua Peng}
\email[]{xhpeng@ustc.edu.cn}
\affiliation{
CAS Key Laboratory of Microscale Magnetic Resonance and School of Physical Sciences, University of Science and Technology of China, Hefei, Anhui 230026, China}
\affiliation{
CAS Center for Excellence in Quantum Information and Quantum Physics, University of Science and Technology of China, Hefei, Anhui 230026, China}

\author{Jiangfeng Du}
\affiliation{
CAS Key Laboratory of Microscale Magnetic Resonance and School of Physical Sciences, University of Science and Technology of China, Hefei, Anhui 230026, China}
\affiliation{
CAS Center for Excellence in Quantum Information and Quantum Physics, University of Science and Technology of China, Hefei, Anhui 230026, China}
\affiliation{Institute of Quantum Sensing and School of Physics, Zhejiang University, Hangzhou 310027, China}

\begin{abstract}
Numerous theories have postulated the existence of exotic spin-dependent interactions beyond the Standard Model of particle physics.
Spin-based quantum sensors, which utilize the quantum properties of spins to enhance measurement precision,
emerge as powerful tools for probing these exotic interactions.
These sensors encompass a wide range of technologies, such as optically pumped magnetometers, atomic comagnetometers, spin masers, nuclear magnetic resonance, spin amplifiers, and nitrogen-vacancy centers.
These technologies stand out for their ultrahigh sensitivity, compact tabletop design, and cost-effectiveness,
offering complementary approaches to the large-scale particle colliders and astrophysical observations.
This article reviews the underlying physical principles of various spin sensors and highlights the recent theoretical and experimental progress in the searches for exotic spin-dependent interactions with these quantum sensors.
Investigations covered include the exotic interactions of spins with ultralight dark matter,
exotic spin-dependent forces,
electric dipole moment,
spin-gravity interactions,
and among others.
Ongoing and forthcoming experiments using advanced spin-based sensors to investigate exotic spin-dependent interactions are discussed.
\end{abstract}


\maketitle


\section{Introduction}


The Standard Model (SM) of particle physics, having undergone extensive testing across various experimental frameworks over many years, stands as a remarkably precise model for explaining subatomic particle behavior\,\cite{particle2022review}.
Despite its considerable success, certain mysteries persist beyond its explanatory capacity.
Notably, the nature of dark matter\,\cite{bertone2005particle} and the strong charge-parity (CP) problem\,\cite{Peccei:1977hh} remain elusive.
The existence of dark matter is supported by substantial gravitational evidence affecting galaxies, yet its fundamental nature remains inexplicably outside the SM's explanatory scope.
There is a variety
of particle candidates for dark matter,
such as weakly interacting massive
particles (WIMPs)\,\cite{aprile2017first,liu2017current,dai2022exotic}, axions and axion-like particles (ALPs)\,\cite{svrcek2006axions,Arvanitaki:2009fg}, and dark photons\,\cite{Abel:2008ai,Goodsell:2009xc}.
The strong CP problem, a theoretical dilemma within quantum chromodynamics (QCD), arises from the exceptionally small value of the neutron's electric dipole moment (EDM)\,\cite{Abel:2020pzs}, a phenomenon that challenges conventional understanding without resorting to finely-tuned parameters.

A concise and elegant solution to both of these perplexing issues emerges with the introduction of a pseudo-scalar particle known as the QCD axion\,\cite{Peccei:1977hh}.
Its inherent potential at its minimum naturally explains the observed absence of the neutron EDM.
Moreover, it offers a plausible explanation for the dark matter relic density through misalignment production\,\cite{preskill1983cosmology, Abbott:1982af, Dine:1982ah}.
{With a sub-electron volt (eV) mass, the axion exhibits a high occupation number when it constitutes the majority of dark matter. As a result, its wavefunction behaves like a coherently oscillating wave over the correlation time and within the de Broglie wavelength. This coherent oscillation can manifest in observables that couple linearly to the axion field.
}
In addition to the QCD axion, fundamental theories featuring extra dimensions, such as string theory, predict the existence of various ultralight bosons\,\cite{svrcek2006axions,Abel:2008ai,Arvanitaki:2009fg,Goodsell:2009xc}.
These encompass pseudo-scalars, known as axion-like particles, and supplemental spin-1 particles, identified as dark photons\,\cite{Abel:2008ai,Goodsell:2009xc}.

Historically,  stellar phenomena, including supernovae, have acted as powerful crucibles for the generation of weakly interacting particles, such as neutrinos.
{Hypothetical particles, such as axions and other light particles, can be efficiently produced through nuclear reactions or thermal processes in the cores of stars if their couplings to SM sectors fall within certain limits\,\cite{raffelt2008astrophysical,arias2012wispy}.}
This ``energy-loss argument" has been widely applied to limit the properties of these newly hypothesized particles, effectively imposing constraints on their characteristics.
Nonetheless, the complexity inherent in astrophysical processes introduces uncertainties or depends on specific models, leading to the possibility that astrophysical limits might significantly fluctuate depending on the analytical methodology employed\,\cite{carenza2019improved,buschmann2022upper,bar2020there,beznogov2018constraints}.

Due to their pseudo-scalar characteristics,
the typical interactions between QCD axions or ALPs and SM fermions manifest through the latter's spin operators.
Even in scenarios where axions or dark photons do not constitute the majority of dark matter, they can still mediate an additional force between SM fermions, governed by spin-dependent potentials\,\cite{Dobrescu:2006au, Fadeev2019}.
These potentials have been recently extensively investigated\,\cite{zhang2023search,Wu2023,wu2022experimental,su2021search,wei2022constraints,feng2022search,aggarwal2022characterization,jiao2021experimental,lee2018improved,kim2019experimental,rong2018constraints,xiao2023femtotesla,xiao2024exotic,wineland1991search,youdin1996limits,kimball2017constraints,chu2013laboratory,bulatowicz2013laboratory,almasi2020new,vasilakis2009limits,glenday2008limits,ji2023constraints,rong2018searching,wang2022limits,wang2023search,ji2018new,kotler2015constraints,hunter2014using,hunter2013using}.
Another intriguing challenge within the SM is the puzzle of the baryon number imbalance existing in the present-day universe.
{Resolving this issue requires a level of CP violation beyond what is predicted by the SM~\cite{Sakharov:1967dj},}
which can be linked to finite EDMs of elementary fermions.
Therefore, the investigation of finite static EDMs shows potential in elucidating the genesis of matter in the cosmos\,\cite{roussy2023improved}.
More generally, the landscape of physics beyond the Standard Model (BSM) comprises numerous exotic interactions between the spins of fermions and hidden sectors,
such as spin-dark matter interactions, and spin-gravity interactions\,\cite{leitner1964parity}.
Fundamental symmetries such as charge-parity-time (CPT) and Lorentz symmetry can also be tested due to the vectorial nature of spins.
These exotic spin-dependent interactions can induce subtle energy shifts of spins,
which may be accessible to laboratory spin-based experiments.

\begin{figure}[b]
    \centering
    \includegraphics[width=0.49\textwidth]{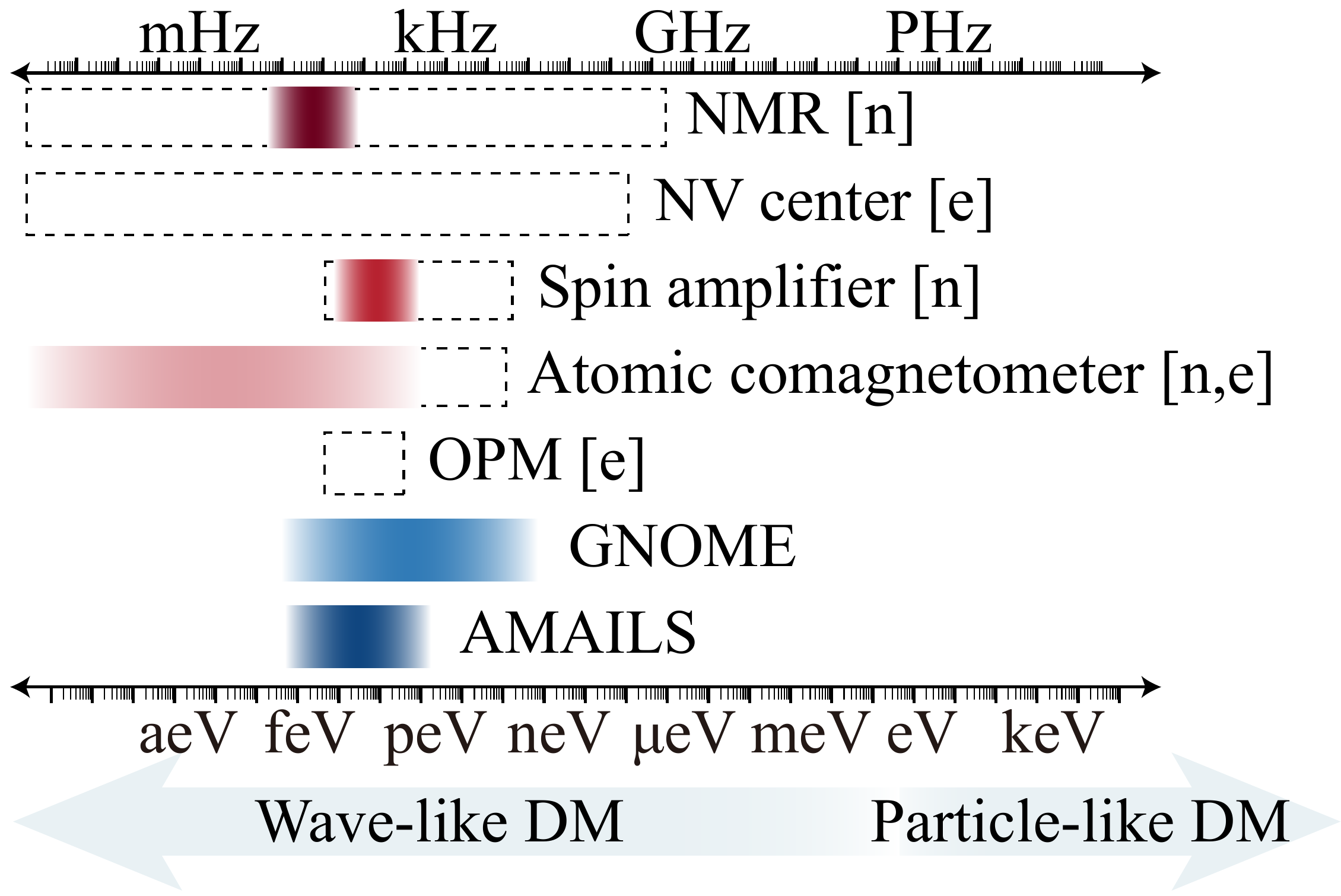}
    \caption{A summary of various spin-based detection techniques for wave-like dark matter (DM). It highlights the application of various methods in probing DM-nucleon interactions, such as nuclear magnetic resonance (NMR)\,\cite{budker2014proposal,garcon2019constraints,wu2019search}, spin amplifiers\,\cite{jiang2021search} and atomic comagnetometers\,\cite{lee2023laboratory,wei2023dark}. Additionally, techniques like nitrogen-vacancy (NV) centers\,\cite{Chigusa:2020gfs}, atomic comagnetometers\,\cite{bloch2020axion} and optically pumped magnetometers (OPMs)\,\cite{Afach:2021pfd} are emphasized for their effectiveness in detecting DM-electron interactions. The GNOME and AMAILS projects, which employ networks of OPMs, focus on different candidates of dark matter: GNOME targets the domain walls of axion-like particles (ALPs)\,\cite{Afach:2021pfd}, while AMAILS aims at dark photons\,\cite{jiang2023search}. The dashed line box indicates the potential DM mass range achievable by the corresponding spin sensor, whereas the color-filled box shows the mass range already achieved by these sensors. 
    {The square brackets [n] and [e] represent the neutron and electron.}
    }
    \label{Figure1}
\end{figure}


Consequently,
spin-based quantum sensors have emerged as promising tools in the pursuit of investigating BSM physics.
For example, Figure\,\ref{Figure1} showcases a range of spin sensor designs, either proposed or already implemented, aimed at detecting wave-like dark matter\,\cite{lee2023laboratory,jiang2021search,wu2019search,garcon2019constraints,bloch2020axion,bloch2022new,bloch2023constraints,wei2023dark,Afach:2021pfd}.
These sensors, known for their tabletop scale and cost-effectiveness, offer a valuable complement to the larger-scale efforts of particle colliders and astrophysical observations.
Moreover, precision experiments designed to detect discrete-symmetry-violating permanent EDMs\,\cite{ellis1989theory,chupp2015electric,kuchler2019searches,chupp2019electric,roussy2023improved,acme2018improved}, exotic spin-dependent interactions mediated by new light bosons\,\cite{zhang2023search,Wu2023,wu2022experimental,su2021search,wei2022constraints,feng2022search,aggarwal2022characterization,jiao2021experimental,lee2018improved,kim2019experimental,rong2018constraints,xiao2023femtotesla,xiao2024exotic,wineland1991search,youdin1996limits,kimball2017constraints,chu2013laboratory,bulatowicz2013laboratory,almasi2020new,vasilakis2009limits,glenday2008limits,ji2023constraints,rong2018searching,wang2022limits,wang2023search,ji2018new,kotler2015constraints,hunter2014using,hunter2013using}, and spin-dependent couplings to ultralight bosonic dark matter fields\,\cite{lee2023laboratory,jiang2021search,wu2019search,garcon2019constraints,bloch2020axion,bloch2022new,bloch2023constraints,wei2023dark,Afach:2021pfd} are pushing the boundaries of our understanding, probing new physics at energy scales far beyond those accessible with current particle colliders.
For example, EDM searches are sensitive to CP-violation due to virtual particles with masses $\gtrsim 10\,\mathrm{TeV}$\,\cite{roussy2023improved,acme2018improved}.
Precision magnetic resonance searches for ALPs are sensitive to phenomena resulting from spontaneous symmetry breaking at energy scales up to those predicted by grand-unification theories and even the Planck scale\,\cite{budker2014proposal}.
Consequently, such tabletop-scale spin-based experiments provide complementary tools to traditional high-energy physics.


With recent advances in quantum technology,
the ability to manipulate quantum states of spins has been significantly improved,
increasing the sensitivity of spin sensors\,\cite{degen2017quantum}.
For example,
in the implementation of electron spins, a thermal atomic vapor serving as a spin sensor for magnetic fields has realized a sensitivity below $1\,\mathrm{fT} / \mathrm{Hz}^{1/2}$\,\cite{kominis2003subfemtotesla,dang2010ultrahigh}, placing them on par with superconducting quantum interference devices (SQUIDs) as the most sensitive magnetometers to date.
Nuclear-spin sensors,
including atomic comagnetometers\,\cite{kornack2005nuclear,smiciklas2011new,vasilakis2009limits,bloch2023constraints,wei2023ultrasensitive} and spin amplifiers\,\cite{jiang2021search,su2021search,wang2022limits}, have also achieved subfemtotesla-level sensitivity.
Despite this, spin sensors still possess significant potential to enhance their sensitivity to the standard quantum limit of approximately $<10\, \mathrm{aT} / \mathrm{Hz}^{1/2}$\,\cite{kominis2003subfemtotesla,allred2002high}.
By employing the entanglement between the spins, spin sensors could potentially achieve unprecedented levels of measurement precision, surpassing even the standard quantum limit\,\cite{degen2017quantum,mitchell2020colloquium}.
Thanks to their exceptional magnetic sensitivity, spin sensors can detect exceedingly subtle energy shifts at scales of approximately $\sim 10^{-26}\,\mathrm{eV}$ induced by exotic interactions\,\cite{graner2016reduced}.
In addition to exploring new physics,
the development of more sensitive spin sensors is broadly driven by other subjects,
including biomagnetic field detection\,\cite{boto2018moving}, inertial rotation measurements\,\cite{kornack2005nuclear}, and chemical analysis\,\cite{theis2011parahydrogen,jiang2018experimental,jiang2019magnetic}.





In recent years, significant advancements have been achieved in the exploration of exotic spin-dependent interactions through the use of spin sensors. This article provides a comprehensive review of the latest progress in the search for such interactions, highlighting the pivotal role of spin sensors in these discoveries.
It begins by outlining the challenges within the SM of particle physics and the presence of exotic spin-dependent phenomena outside the realm of the SM (see Sec.\,\ref{SecII}).
Then various spin sensors are discussed in Sec.\,\ref{Sec:sensor},
including optically pumped magnetometers (OPMs), atomic comagnetometers, spin masers, spin amplifiers, nuclear magnetic resonance (NMR), nitrogen-vacancy in diamonds, and others.
Moreover, the article discusses precision measurements of exotic spin-dependent interactions using these sensors in Sec.\,\ref{SecIV},
including searches for ultralight bosonic dark matter,
exotic spin-dependent forces,
EDMs,
spin-gravity interactions,
and others.
Networks of distributed spin sensors are presented, along with a summary of their applications in the pursuit of exotic spin-dependent interactions (see Sec.\,\ref{SecV}).
Ongoing and future experiments of the next decade are discussed in Sec.\,\ref{SecVI}.

\section{Overview of exotic spin-dependent interactions and detectable effects}
\label{SecII}

\begin{figure}[t]
    \centering
    \centering
    \includegraphics[width=0.47\textwidth]{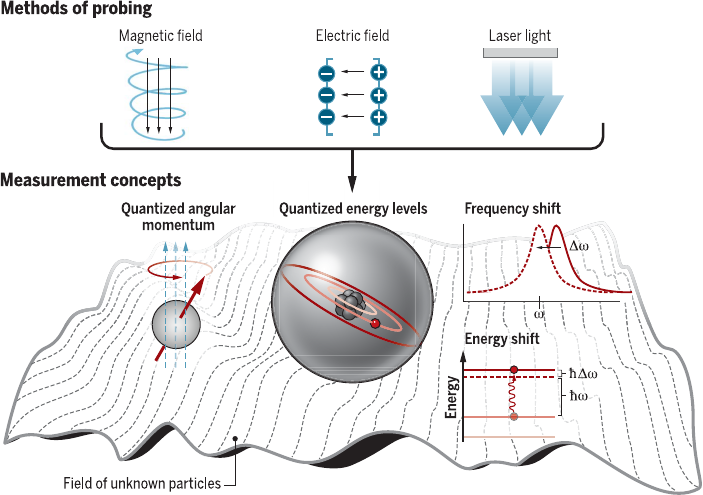}
    \caption{The principle of searching for exotic spin-dependent interactions with atoms and molecules, where spin sensors represent a crucial category of detectors for exotic interactions. These sensors operate by measuring the subtle energy shifts in the spins induced by the new physics field. This figure is from Ref.\,\cite{demille2017probing}.}
    \label{Figure2}
\end{figure}

Spin sensors find diverse applications in the search for dark matter\,\cite{lee2023laboratory,jiang2021search,wu2019search,garcon2019constraints,bloch2020axion,bloch2022new,bloch2023constraints,wei2023dark,Afach:2021pfd}, spin-dependent exotic forces\,\cite{zhang2023search,Wu2023,wu2022experimental,su2021search,wei2022constraints,feng2022search,aggarwal2022characterization,jiao2021experimental,lee2018improved,kim2019experimental,rong2018constraints,xiao2023femtotesla,xiao2024exotic,wineland1991search,youdin1996limits,kimball2017constraints,chu2013laboratory,bulatowicz2013laboratory,almasi2020new,vasilakis2009limits,glenday2008limits,ji2023constraints,rong2018searching,wang2022limits,wang2023search,ji2018new,kotler2015constraints,hunter2014using,hunter2013using}, precision measurements of particle properties\,\cite{roussy2023improved,acme2018improved}, and the testing of fundamental principles in physics\,\cite{Bluhm:1999ev,Brown:2010dt,Smiciklas:2011xq}.
The exotic interaction between a spin sensor and a new physics field, as depicted in Fig.\,\ref{Figure2}, can lead to minor energy shifts in the spins.
This interaction can be quantified by the Hamiltonian:
\begin{equation}
H= \alpha \, \vec{\mathcal{O}}\cdot \vec{\sigma},
\label{hamD}
\end{equation}
where $\alpha$ represents the strength of the exotic interaction,
$\vec{\sigma}$ denotes the spin projection vector of SM fermions within the spin sensor, and $\vec{\mathcal{O}}$ represents the new physics field. This field and its implications are further explored in subsequent sections. 
The exotic interaction described by Eq.\,\eqref{hamD} is similar to Zeeman interactions, potentially inducing a pseudomagnetic field on the spins. Such a pseudomagnetic field could be observable with highly sensitive spin sensors.

The Hamiltonian described in Eq.\,\eqref{hamD} serves as a powerful tool for exploring a variety of physical phenomena. For instance, ultralight bosonic fields can generate time-varying signal in $\vec{\mathcal{O}}$ with frequency that typically match the energy of the bosons\,\cite{graham2013new,budker2014proposal}. Additionally, exotic spin-dependent forces may involve interactions between two vertices, where one is characterized by Eq.\,(\ref{hamD}) and the other could be either monopolar or dipolar\,\cite{Dobrescu:2006au,Fadeev2019}. The detection of EDMs is another case, requiring the application of an electric field to $\vec{\mathcal{O}}$ and the measurement of a static shift in the coupling parameter $\alpha$\,\cite{chupp2019electric}. Furthermore, in testing spin-gravity interactions, the Earth's gravitational acceleration vector $\vec{g}$ is incorporated into $\vec{\mathcal{O}}$\,\cite{venema1992search}.

\subsection{Ultralight bosonic dark matter}
\label{interaction}



Ultralight bosons, characterized by their extraordinarily low masses in the sub-eV range, have recently garnered significant attention as potential candidates for natural dark matter\,\cite{budker2014proposal,kimball2023probing}. 
{When bosons constitute the majority of dark matter, their wavefunctions behave like coherently oscillating waves~\cite{Hu:2000ke}, driven by their high occupation number in the sub-eV mass range. This results in the periodic oscillation of the background bosonic field $\phi$, expressed as $\phi \propto \cos(\omega_\phi t)$, where $\omega_{\phi}$ represents the characteristic frequency.}
In scenarios involving cold dark matter, the frequency $\omega_{\phi}$ closely matches the mass of the non-relativistic bosonic field, with a minor velocity dispersion $\sim 10^{-3}$ of the speed of light arising from the virialization process during galaxy halo formation, where gravitational potential energy is converted into kinetic energy. This translates to a narrow bandwidth, typically around $\sim 10^{-6}$ of the central frequency\,\cite{budker2014proposal,graham2013new}. According to the standard halo model for wave-like bosonic dark matter, the velocity distribution around Earth is isotropic in the galactic frame, resulting in a relative velocity on the order of $10^{-3}$ of the speed of light with respect to terrestrial detectors, primarily due to the Earth's rotation\,\cite{Foster:2017hbq,Derevianko:2016vpm}. The local density of this standard halo model is approximately $0.45$\,GeV/cm$^3$.

What enhances the intrigue surrounding ultralight bosonic dark matter is its capacity to form localized compact structures, leading to significantly increased density when aggregated\,\cite{Banerjee:2019epw,Banerjee:2019xuy,Kim:2021yyo}, thereby substantially enhancing detection prospects. Additionally, the self-interactions among these bosons can result in exponential growth of field values\,\cite{Budker:2023sex}.

Beyond these ultralight bosonic dark matter scenarios, various other forms of bosonic field backgrounds may exist, even if they do not constitute the majority of dark matter. These include cosmological relativistic backgrounds\,\cite{Baumann:2016wac, Dror:2021nyr,Chen:2024aqf}, which can have spectra dependent on their production mechanisms and the thermal history of the universe. Moreover, one can anticipate the presence of incoming bosonic waves originating from specific directions towards Earth, such as a cold stream component of dark matter\,\cite{OHare:2017yze, Foster:2017hbq, Knirck:2018knd}, as well as waves resulting from transient astrophysical phenomena\,\cite{dailey2020quantum,Arakawa:2023gyq}. The latter category may include dipole radiations from binary systems charged by hidden $U(1)$ forces\,\cite{Krause:1994ar,Dror:2021wrl,Hou:2021suj}, and emissions from strongly self-interacting gravitational atoms\,\cite{baryakhtar2021black,East:2022ppo}.

Lastly, ultralight bosonic fields can give rise to topological defects in the early universe, characterized by spatially varying values of the bosonic field. These defects can produce transient signals when they traverse terrestrial sensors\,\cite{pospelov2013detecting,Derevianko:2013oaa,Afach:2021pfd}, often necessitating a network of global detectors, such as the Global Network of Optical Magnetometers for Exotic physics (GNOME)\,\cite{pospelov2013detecting, Pustelny:2013rza, Afach:2018eze, Afach:2021pfd}, to reconstruct their trajectories and properties.


Spin sensors play a pivotal role in the search for various couplings between ultralight bosons and SM fields. These couplings essentially translate the time-varying boson wave functions into corresponding signals detectable by these sensors. In this review, our primary focus will be on two types of bosons: spin-$0$ pseudoscalar particles\,\cite{lee2023laboratory,jiang2021search,wu2019search,garcon2019constraints,bloch2020axion,bloch2022new,bloch2023constraints,wei2023dark,Afach:2021pfd}, such as axions, and spin-$1$ vector particles\,\cite{jiang2023enhanced}, like dark photons. Both axions and dark photons arise as generic predictions from fundamental theories with extra dimensions\,\cite{svrcek2006axions,Abel:2008ai,Arvanitaki:2009fg,Goodsell:2009xc}.

A prominent example is the QCD axion, serving dual purposes as a solution to the strong CP problem by explaining the vanishing electric dipole moment\,\cite{Peccei:1977hh} and as a viable candidate for dark matter\,\cite{preskill1983cosmology, Abbott:1982af, Dine:1982ah}. One of the typical couplings associated with the QCD axion is the axion-gluon coupling, which leads to intriguing phenomena like the oscillating neutron EDM in the presence of a background QCD axion\,\cite{graham2013new,budker2014proposal}. This coupling necessitates the existence of an electric field within nuclear or atomic systems, capable of interacting with the oscillating EDM\,\cite{aybas2021search}.

Another intriguing coupling that necessitates the use of spin sensors is the axion-fermion interaction. In the non-relativistic limit of the fermion field, this interaction gives rise to the coupling described by Eq.\,(\ref{hamD}), which links the spin of fermions to the spatial gradient of axions:
\be g_{a\psi} a \bar{\psi} \gamma_5 \psi \rightarrow \vec{\mathcal{O}}_a = \vec{\nabla} a.  \label{Lae}\ee
Here $a$ represents the axion field, $\psi$ is the fermion field, $g_{a\psi}$ denotes the coupling constant, and $\gamma^5\equiv i\gamma^0\gamma^1\gamma^2\gamma^3$ represents the product of gamma matrices $\gamma^\mu$.
Consequently, the gradient of axions effectively acts as a pseudomagnetic field, resulting in the precession of the spins of initially polarized fermions\,\cite{graham2013new,budker2014proposal,stadnik2014axion,Barbieri:2016vwg,abel2017search,Stadnik:2017mid,jiang2021floquet,wu2019search,garcon2019constraints,smorra2019direct,Mitridate:2020kly,Chigusa:2020gfs,aybas2021search,jiang2021search,Kim:2021eye,Poddar:2021ose}.
This unique phenomenon has been extensively explored in the literature, leading to a multitude of experimental efforts aimed at detecting axion-nucleon interactions. These detection strategies include nuclear magnetic resonance experiments\,\cite{graham2013new,budker2014proposal,garcon2019constraints,aybas2021search}, Floquet maser techniques\,\cite{jiang2021floquet}, and the utilization of spin amplifiers\,\cite{jiang2021search,su2021search}. Additionally, there are detection methods based on axion-electron couplings, such as axion-magnon conversion\,\cite{Barbieri:2016vwg,Mitridate:2020kly,Chigusa:2020gfs}.

Expanding our scope beyond axions, we consider the coupling between dark photons ($V_\mu$) and fermion spins. For operators up to dimension-5, relevant interactions include axial-vector (A), EDM, and magnetic dipole moment (MDM)-type couplings\,\cite{graham2018spin,Chen:2021bdr}:
    \begin{align}
    g_A V_{\mu} \bar{\psi}\gamma^{\mu} \gamma^5 \psi &\rightarrow \vec{\mathcal{O}}_A=\vec{V}, \label{axialvec}\\
    g_{\text{MDM}} V_{\mu \nu} \bar{\psi} \sigma^{\mu \nu} \psi &\rightarrow \vec{\mathcal{O}}_{\text{MDM}}=\vec{\nabla}\times \vec{V},\label{MDMLH}\\
      g_{\text{EDM}} V_{\mu \nu} \bar{\psi}\sigma^{\mu \nu} i \gamma^5 \psi &\rightarrow \vec{\mathcal{O}}_{\text{EDM}}=\partial_0 \vec{V} - \vec{\nabla} V^0.\label{EDMLH}
\end{align}
Here $g_A$, $g_{\text{EDM}}$, and $g_{\text{MDM}}$ are the corresponding coupling constants, respectively. The anti-symmetric tensor $\sigma^{\mu \nu}\equiv\gamma^\mu\gamma^\nu-\gamma^\nu\gamma^\mu$ is constructed from Dirac matrices, while $V_{\mu\nu}\equiv \partial_\mu V_\nu -\partial_\nu V_\mu$ represents the field strength tensor of the dark photon. To distinguish between different types of bosonic dark matter and other sources, a network of spin sensors is employed, with the manipulation of sensor baseline and orientation providing crucial information\,\cite{Chen:2021bdr}.

Furthermore, a massive dark photon can kinetically mix with the SM photon through a dimension-$4$ operator $\varepsilon F_{\mu\nu} F^{\prime \mu\nu}$, where $\varepsilon$ is the kinetic mixing coefficient, and $F_{\mu\nu}$ and $F^{\prime}_{\mu\nu}$ are the field stress tensors of the SM photon and dark photon, respectively. In the interaction basis, the dark photon field acts as an effective electric current\,\cite{Chaudhuri:2014dla}, capable of exciting electromagnetic fields within shielded regions. Magnetometers can be employed to capture the magnetic fields induced by this effective current. Notably, the magnitude of the induced magnetic fields is often proportional to the size of the shielded region. Strategies to enhance signal detection using large shielded regions include leveraging geomagnetic fields\,\cite{Fedderke:2021aqo,Fedderke:2021rrm} and employing large shielded rooms equipped with a network of spin sensors\,\cite{jiang2023search}.




\subsection{Exotic spin-dependent forces}
\label{forcesearches}

Spin sensors find intriguing applications in the quest for exotic spin-dependent forces, which are experiments designed to uncover potential interactions between SM fermions mediated by the presence of new bosons\,\cite{Moody:1984ba,Dobrescu:2006au,Fadeev2019}. These interactions involve a product of two vertices, with one vertex being dipolar, taking the form of Eq.\,(\ref{hamD}). The other vertex can either be monopolar or dipolar. The simplest monopolar vertex can result from a Yukawa-like coupling mediated by a scalar or a gauge coupling mediated by a vector between the SM fermions and new bosons. In contrast, the dipolar coupling arises from a pseudo-scalar vertex as described in Eq.\,(\ref{Lae}) or from dark photons with non-minimal couplings.
It is worth noting that these fifth-force potentials, being mediated by virtual bosons,
do not necessarily dominate the dark matter content.

A notable example is the exotic spin-spin interaction mediated by a QCD axion\,\cite{Moody:1984ba},
which can couple to nucleons through the pseudo-scalar portal.
The potential associated with this interaction is\,\cite{Moody:1984ba,Dobrescu:2006au,Fadeev2019} 
\begin{equation}
\begin{aligned}
V_3 & = \frac{g_p g_p \hbar^3}{16 \pi m_1 m_2 c}\left[\left(\vec{\sigma}_1 \cdot \vec{\sigma}_2\right)\left(\frac{1}{\lambda r^2}+\frac{1}{r^3}\right)\right. \\
& \left.-\left(\vec{\sigma}_1 \cdot \hat{r}\right)\left(\vec{\sigma}_2 \cdot \hat{{r}}\right)\left(\frac{1}{\lambda^2 r}+\frac{3}{\lambda r^2}+\frac{3}{r^3}\right)\right] e^{-r / \lambda},
\end{aligned}
\end{equation}
where $g_{p}g_{p}$ is the pseudoscalar coupling constant, $m_{1,2}$ is the mass of two interacting fermions, $c$ is the speed of light, $\vec{\sigma}_{1,2}$ is the spin vector,
$r$ is the distance between the two interacting spins,
and $\hat{{r}}$ is the corresponding unit vector.
Here $\lambda \!= \!\hbar(m_b c)^{-1}$ is the force range (or the boson Compton wavelength) with $m_b$ being the mass of the boson mediator.

In a broader context,
relying on rotational invariance in the center-of-mass frame of two SM fermions, there exist a total of $16$ independent operators\,\cite{Dobrescu:2006au} that can give rise to distinct fifth-force potentials.
With the exception of the monopole-monopole interaction,
all the other $15$ are dependent on spins.
Thus, spin sensors offer promising tools to investigate such exotic spin-dependent forces.
Some of these operators involve relative velocities between two test fermions, necessitating dynamic motion between spin sensors and other test fermions. Detailed relations between the fifth-force potentials and the interaction Lagrangian of the two vertices can be found in Refs.\,\cite{Dobrescu:2006au,Fadeev2019,Fadeev:2019jzi,Costantino:2019ixl}.

Typically, the fifth-force potential adopts the form $e^{-r/\lambda}$. Consequently, the Compton wavelength of the new boson dictates the range of the force. As constraints become more stringent for higher boson masses, spin sensors become indispensable for probing smaller scales. Quantum forces between SM fermions can emerge through the exchange of multiple new mediators\,\cite{Brax:2017xho,Costantino:2019ixl}. These induced potentials generally exhibit shorter force ranges compared to those arising from single-mediator interactions.


\subsection{Electric dipole moments}


Electric Dipole Moments (EDMs) are fundamental properties of particles,
representing the coefficient of linear response of their angular momentum when these particles are subjected to an external electric field.
These quantities bear a profound connection to the phenomenon of CP violation, offering a gateway to explore the delicate intricacies of symmetry breaking in the universe\,\cite{chupp2019electric,cairncross2019atoms}.Deviations from the expected EDM values predicted by the SM have the potential to uncover new physics, casting light upon the elusive mechanisms underlying CP violation.
{To measure EDMs, an external electric field is introduced, the corresponding Hamiltonian is\,\cite{chupp2019electric,cairncross2019atoms}
\begin{equation}
    H_{\textrm{edm}}=\vec{\sigma}\cdot \vec{B}+\vec{d} \cdot \vec{E},
\end{equation}
where $\vec{\sigma}$ is the spin vector, $\vec{B}$ is the magnetic field, $\vec{d}$ is the EDM of fundamental particles, atoms or molecules along the total angular momentum of a particular state, and $\vec{E}$ is the external electric field.
The ultimate aim of these measurements is to precisely determine the EDMs $d$.}

Within the framework of the SM, which primarily accounts for CP violation through weak interactions, EDMs are predicted to be astonishingly minuscule. For example, the SM foresees EDMs for quarks\,\cite{Shabalin:1978rs,Shabalin:1980tf,Shabalin:1982sg,Eeg:1982qm,Eeg:1983mt,Khriplovich:1985jr,Czarnecki:1997bu} emerging at the three-loop level of Feynman diagrams, while the electron's EDM is even smaller\,\cite{Pospelov:1991zt,Booth:1993af,Pospelov:2013sca,Yamaguchi:2020eub}. The remarkable tininess of these predictions accentuates their sensitivity to any physics beyond the SM, underscoring the pivotal role played by EDM measurements in the exploration of novel physics phenomena\,\cite{Pospelov:2005pr}.

Atomic and molecular EDMs are inherently intertwined with the constituents and their distribution. This dependence arises due to the ability of internal electrons to screen the applied electric field through rearrangement, a principle encapsulated within Schiff's theorem\,\cite{Schiff:1963zz}. Furthermore, the remaining EDMs are intricately linked to the concept of Schiff moments\,\cite{Flambaum:2001gq}, which arises from the distribution of charge and currents within nuclei. 

\begin{table}[b]
\newcommand{\tabincell}[2]{\begin{tabular}{@{}#1@{}}#2\end{tabular}}
\begin{ruledtabular}
\caption {Summary of published parameters about diamagnetic atoms\,\cite{chupp2019electric}. } 
\label{tab:my_label2}
\renewcommand{\arraystretch}{1.2}
\begin{tabular}{c c c}   
System &  ${ }^{129} \mathrm{Xe}$&${ }^{199} \mathrm{Hg}$\\
\midrule
$k_S\,(\mathrm{~cm/fm^{3}})$&$0.27 \times 10^{-17}$&$-2.8 \times 10^{-17}$\\[0.15cm]
$a_n$\,(fm$^{2}$) &0.63&1.9\\[0.15cm]
$a_p$\,(fm$^{2}$)&0.125&0.2\\[0.15cm]
$a_0$\,($e\,\mathrm{fm}^3$)&-0.008&0.01\\[0.15cm]
$a_1$\,($e\,\mathrm{fm}^3$)&-0.006 &$\pm 0.02$\\[0.05cm]
\end{tabular} 
\end{ruledtabular}
\label{table_EDM}
\end{table}

Contributions to the EDMs in systems can be described using a specific set of low-energy parameters\,\cite{chupp2019electric,kuchler2019searches,chupp2015electric,seng2014nucleon}.
Taking diamagnetic atoms as an example, their atomic EDMs can be expressed as follows
\begin{equation}
d_A=k_S S-\left[k_T^{(0)} C_T^{(0)}+k_T^{(1)} C_T^{(1)}\right]+\cdots,
\end{equation}
where $S$ represents the Schiff moment, ${C_T}^{(0,1)}$ denote the isoscalar and isovector electron-quark tensor interactions, respectively, and $k_S$, $k_{T}^{(0,1)}$ are the sensitivity coefficients for these parameters\,\cite{chupp2019electric,fleig2018model,ginges2004violations}.
The Schiff moment includes contributions from unpaired nucleons ($d_n,d_p$) and the long-range pion-nucleon coupling $\bar{g}_\pi^{(0,1)}$
\begin{equation}
    S=a_{\mathrm{p}} d_{\mathrm{p}}+a_{\mathrm{n}} d_{\mathrm{n}}+\frac{m_N g_A} {F_\pi}[a_0 \bar{g}_\pi^{(0)}+a_1 \bar{g}_\pi^{(1)}]
\end{equation}
where the factor $m_N g_A / F_\pi \sim 13.5$ and the coefficients $a_{\mathrm{p}}, a_{\mathrm{n}}, a_0, a_1$ are specific to particular diamagnetic atoms. Detailed values and discussions of these parameters are provided in\,\cite{chupp2019electric,skripnikov2017enhanced,dzuba2011relations,ginges2004violations,fleig2018model}.
Table\,\ref{tab:my_label2} shows $k_S$ and the Schiff moment related parameters $(a_n, a_p, a_0, a_1)$ for $^{129}$Xe and $^{129}$Hg atoms.


\subsection{Spin-gravity interactions}

An intriguing illustration of spin-gravity interaction is found in the interplay between the intrinsic spins of fermions and the gravitational field vector\,\cite{leitner1964parity,mcreynolds1951gravitational,dabbs1965gravitational,flambaum2009scalar}, denoted as $\vec{g}$ and symbolized by the vector field $\vec{\mathcal{O}}$ in Eq.\,(\ref{hamD})
\begin{equation}
\label{SG}
    H_{\textrm{sg}}=\chi \vec{\sigma} \cdot \vec{g}({r}),
\end{equation}
where $\chi$ is the particle’s gyrogravitational ratio, $\vec{g}({r})$ represents the acceleration due to gravity, and $r$ is the distance from the geographical centre of Earth.
Therefore, the acceleration of spins in gravity depends on the their Zeeman sublevels.
The energy difference induced by the spin-gravity interactions between the adjacent energy levels is $\hbar A_{\rm{sg}}(r)=-\hbar \chi g(r) $.
The difference in the acceleration is $\left|\delta g_s\left(r\right)\right|=2 \hbar\left|A_{\mathrm{sg}}\right| / (m_{n,e} r)$, where $m_{n,e}$ is the mass of neutron or electron.
The ``Eötvös'' parameter $\eta_{\mathrm{sg}}$ can be further defined as 
\begin{equation}
   \eta_{\mathrm{sg}}=\frac{\left|\delta g_{s, n}\left(r\right)\right|}{g\left(r\right)}=\frac{2 \hbar\left|A_{\mathrm{sg}}\left(r\right)\right|}{m_{n,e} g\left(r\right) r}.
\end{equation}
It is the ratio between the enengy of spin-gravity interactions and the gravitational potential.

\begin{table*}[t]
\footnotesize
\newcommand{\tabincell}[1]{\begin{tabular}{@{}#1@{}}#1\end{tabular}}
\begin{ruledtabular}
\caption {
Specifications for various spin sensors are detailed in the catalog. For optically pumped magnetometers (OPMs), their energy resolution is inferred from their sensitivity data. The sensitivity specified for the K-$^3$He atomic comagnetometer relates to its ability to detect pseudomagnetic fields.
The size refers to the dimensions of the sensing element, such as a vapor cell, utilized in the spin sensor.
The Cosmic Axion Spin Precession Experiment (CASPEr) has set limits on axion-like dark matter, although it does not explicitly report its magnetic sensitivity and energy resolution. This is also true for $^{129}$Xe-$^3$He comagnetometer, which is utilized in searches for the Xenon EDM. Regarding Ramsey interferometry, the noted precision is associated with the statistical uncertainty.} 
\label{tab:my_label1}
\renewcommand{\arraystretch}{1.2}
\begin{tabular}{c c c c c c c}
\multirow{2}{2cm}{\centering Type}  & \multirow{2}{*}{\centering Spin species} & Sensitivity &  Energy resolution  & Size  & \multirow{2}{*}{\centering Ref.}&\multirow{2}{3cm}{\centering Applications}\\
 &  & (or precision) &  $(\mathrm{eV}/{\mathrm{Hz}^{1/2}})$ & (cm) & \\
\midrule
\multirow{3.6}{2cm}{\centering OPM} & K & 0.16\,$\mathrm{fT}/{\mathrm{Hz}^{1/2}}$@40\,Hz & $\sim 3.1 \times 10^{-21}$ & $\sim1$ & \cite{dang2010ultrahigh}& \multirow{3.6}{3cm}{\centering Spin-dependent force search\,\cite{kim2019experimental,xiao2023femtotesla,xiao2024exotic,wu2022experimental,ji2018new,ji2023constraints}, DM search\,\cite{Afach:2021pfd}.}\\[0.15cm]
& Rb & 5\,$\mathrm{fT}/{\mathrm{Hz}^{1/2}}$@50-200\,Hz & $ \sim 9.5\times 10^{-19}$ & $\sim$0.1&\,\cite{griffith2010femtotesla} \\[0.15cm]
& Cs & 40\,$\mathrm{fT}/{\mathrm{Hz}^{1/2}}$@30\,Hz  &$ \sim 5.6\times 10^{-19}$&$\sim2$ &\cite{ledbetter2008spin}\\
\midrule
\multirow{7.7}{2cm}{\centering Atomic comagnetometer} & K-$^{3}\textrm{He}$ &0.75\,$\mathrm{fT}/{\mathrm{Hz}^{1/2}}$@0.18\,Hz &$1.0 \times 10^{-22}$ & $\sim$ 2.4&\cite{vasilakis2009limits}&  \multirow{7.7}{3cm}{\centering DM search\,\cite{wei2023dark,lee2023laboratory,bloch2020axion}, spin-dependent force search\,\cite{wei2022constraints,lee2018improved,bulatowicz2013laboratory,venema1992search,vasilakis2009limits,feng2022search,zhang2023search,venema1992search,kimball2017constraints}, EDMs search\,\cite{abel2020measurement,allmendinger2019measurement}, spin-gravity interaction search, CPT symmetry\,\cite{brown2011new}.} \\[0.15cm]
& $\mathrm{K}$-$\mathrm{Rb}$-${ }^{21}\mathrm{Ne}$ & $
3 \times 10^{-8}\,\mathrm{rad} / \mathrm{s} / \mathrm{Hz}^{1/2}
$@0.2-$1.0 \mathrm{~Hz}$  & $2.1 \times 10^{-23}$ & $\sim$1.2&\cite{wei2023ultrasensitive} \\[0.15cm]
& $^{129}$Xe-$^{131}$Xe & 
$1 \times 10^{-7} \mathrm{~Hz} / \mathrm{hr}^{1 / 2}$@ $\Omega_m$\footnote{$\Omega_m$ represents the frequency of Earth's rotation.}/$(2 \pi)$  & $10^{-20}$ & $\sim$1&\cite{zhang2023search}\\[0.15cm]
& $^{199}$Hg-$^{133}$Cs & $
0.8\,\mu \mathrm{Hz}$ & $-$& $\sim$1 &\cite{peck2012limits}
\\[0.15cm]
& $^{199}$Hg-$^{201}$Hg & $0.5\,\mu \mathrm{Hz}$ &$-$ &$\sim$2 &\cite{venema1992search}
\\[0.15cm]
& $^{129}$Xe-$^{3}$He & $-$ &$-$ &$\sim$2 &\cite{abel2020measurement}\\
\midrule
\multirow{2.35}{2cm}{\centering Spin Amplifier} & Rb-$^{129}\textrm{Xe}$  & 18\,$\mathrm{fT}/{\mathrm{Hz}^{1/2}}$@5-30\,Hz & $8.8 \times 10^{-22}$ & $\sim$1& \cite{jiang2021search} &\multirow{2.35}{3cm}{\centering DM search\,\cite{jiang2021search}, spin-dependent force search\,\cite{su2021search,wang2022limits,wang2023search}.}\\[0.15cm]
& Rb-$^{129}\textrm{Xe}$ (Floquet) & $\sim $20\,$\mathrm{fT}/{\mathrm{Hz}^{1/2}}$@5-30\,Hz  &  $\sim9.8 \times 10^{-22}$ &$\sim$1 &\cite{jiang2022floquet} \\
\midrule
\multirow{3.9}{2cm}{\centering CASPEr NMR}  & ${ }^{207}\mathrm{Pb}$ (Solid)  & $-$&$-$&$\sim 0.5$\,cm&\cite{aybas2021search}  &\multirow{3.8}{3cm}{\centering DM search\,\cite{garcon2019constraints,wu2019search,aybas2021search}.}   \\[0.15cm]
  & \multirow{2.35}{2cm}{\centering ${ }^{13} \mathrm{CH}_3 \mathrm{CN}$, $\mathrm{H} {}^{13}\mathrm{COOH}$ (ZULF)}  & \multirow{2.35}{2cm}{\centering $-$}&\multirow{2.35}{2cm}{\centering$-$} & \multirow{2.35}{2cm}{\centering $\sim 0.5$\,cm}&   \multirow{2.35}{0.6cm}{\centering \cite{wu2019search,garcon2019constraints}} \\[0.6cm]
\midrule
\multirow{3.6}{2cm}{\centering Nitrogen-vacancy (NV) center} &  Single NV&  $0.5~{\mathrm{nT/Hz}^{1/2}}$& $-$ & $\sim3\times10^{-6}$ &  \cite{Zhao2023}&\multirow{3.6}{3cm}{\centering Spin-dependent force search\,\cite{rong2018constraints,rong2018searching,jiao2021experimental,liang2022}.}\\[0.15cm]
&  \multirow{2.35}{2cm}{\centering Ensemble NV} & $0.21~{\mathrm{pT/Hz}^{1/2}}$ (AC)&$-$& $\sim7\times10^{-3}$&\cite{barry2023sensitive}&\\[0.15cm]
&  &  $0.46~{\mathrm{pT/Hz}^{1/2}}$ (DC)& $-$ & $\sim7\times10^{-3}$ &\cite{barry2023sensitive}\\
\midrule
\multirow{3.6}{2cm}{\centering Spin maser} & $^{129}$Xe-$^{131}$Xe & $6.2\,\mu\mathrm{Hz}$&$-$&$\sim$2&\cite{sato2018development} &\multirow{3.6}{3cm}{\centering EDM search\,\cite{sato2018development,inoue2011experimental}, spin-dependent force search\,\cite{glenday2008limits}, CPT symmetry\,\cite{bear2000limit}.}\\[0.15cm]
& $^{129}$Xe-$^{3}$He & $6.1$\,nHz&$-$&$\sim$2&\cite{glenday2008limits}\\[0.15cm]
& $^{129}$Xe-$^{87}$Rb (Floquet) & $700\,$fT/Hz$^{1/2}@1$-$60\,\textrm{mHz}$& $3.4\times 10^{-20}$&$\sim$1&\cite{jiang2021floquet}\\
\midrule
\multirow{2.35}{2cm}{\centering Ramsey Interferometry}&HfF$^{+}$&$ 22.8\,\mu \mathrm{Hz}$&$-$&$-$&\cite{roussy2023improved}&\multirow{2.35}{3cm}{\centering EDM search\,\cite{roussy2023improved,acme2018improved}, spin-dependent force search\,\cite{piegsa2012limits}.}\\[0.15cm]
&ThO& $59.4\,\mu\mathrm{Hz}$ &$-$&$-$&\cite{acme2018improved}
\end{tabular} 
\end{ruledtabular}
\end{table*}

Although the spin-gravity interaction is presently described in a phenomenological manner, it beckons for a quest to establish an ultraviolet completion within the current theoretical framework. What sets this interaction apart is its unique property of violating both parity inversion (P) and time reversal (T) symmetry, rendering it a captivating avenue for delving into fundamental symmetries and corroborating equivalence principles\,\cite{Peres:1977yh,leitner1964parity}.

On the other hand, as predicted by general relativity, a rotating gravitational field exerts an effect known as frame dragging, causing nearby spinning objects to precess\,\cite{1916MNRAS..76..699D,1918PhyZ...19..156L}. This phenomenon implies that the orbital angular momentum and the rotation of the Earth could couple with the spins of fermions\,\cite{Mashhoon:2000jq,Mashhoon:2013jaa}. Detecting and quantifying such intricate spin-gravity interactions represent a formidable challenge, demanding levels of sensitivity beyond the capabilities of current technology\,\cite{Fadeev:2020gjk}.

\subsection{Lorentz symmetry and charge-parity-time violation}

Lorentz symmetry and Charge-Parity-Time (CPT) symmetry are foundational to the SM of particle physics. Within the SM framework, the electroweak interaction disrupts CP symmetry, hinting at the possibility that CPT symmetry could also be vulnerable to breaches at elevated energy levels. This notion, often propelled by specific models of quantum gravity\,\cite{Mavromatos:2005mi}, underscores the importance of thorough examination. Traditional approaches to testing CPT symmetry have typically centered around the utilization of anti-particles as benchmarks for charge conjugation. 
{However, since the derivation of the CPT theorem is specifically valid for Lorentz-invariant local quantum field theories with a Hermitian Hamiltonian~\cite{Schwinger:1951xk, Luders:1954zz}, investigating Lorentz symmetry—without requiring the presence of antiparticles—becomes a natural approach for testing CPT symmetry conservation. In fact, it has been demonstrated that CPT violation necessarily implies a violation of Lorentz invariance~\cite{Greenberg:2002uu}. Nevertheless, the converse is not always true, as some models allow for Lorentz violation without violating CPT symmetry~\cite{Potting:2013pqa}.}
In this vein, employing spin sensors for Lorentz symmetry tests leverages Earth's rotation to detect any orientation-dependent variations in precession rate\,\cite{Bluhm:1999ev,Brown:2010dt,Smiciklas:2011xq}. Such discrepancies could signal the presence of Lorentz-violating interactions, providing a novel avenue for investigating the integrity of these fundamental symmetries.

\subsection{Dirac monopole}

The pursuit of magnetic monopoles is widely recognized as a cutting-edge endeavor in the realms of modern physics and astrophysics. This search predominantly focuses on capturing their unique induction signals and the energy they deposit.
Currently, the most sensitive detection methods employ superconducting coils\,\cite{dusad2019magnetic} and
large deep-underground ultra-low-background detectors\,\cite{aartsen2016searches,acharya2022search}.
Furthermore, the advent of quantum precision measurement technologies has introduced significant benefits in enhancing measurement precision and reducing the cost and size of the necessary instruments.
This innovation facilitates a shift away from the traditional dependency on costly, large-scale scientific equipment and stringent experimental conditions.
The Search for Cosmic Exotic Particles (SCEP) project pioneers a novel coincidence measurement approach that integrates room-temperature spin sensors with plastic scintillators for the detection of magnetic monopoles\,\cite{SCEP2024}.
{The room-temperature induction coils and plastic scintillators allow for the collection of both the induction and scintillation signals generated by the passage of a magnetic monopoles. 
These signals are detected by the radio-frequency atomic magnetometers with a high sensitivity of several fT/Hz$^{1/2}$\,\cite{kominis2003subfemtotesla,savukov2005tunable}.
This approach provides acceptance to the magnetic monopoles with their velocities larger than about $10^{6}$ light speed and their masses larger than approximately $10^7$\,GeV.
The exceptional sensitivity of spin sensors has the potential to significantly boost the signal-to-noise ratio in magnetic monopole searches, heralding a new phase in the detection and study of these elusive particles.}

\section{Overview of spin-based quantum sensors}
\label{Sec:sensor}

Spin is intricately linked to magnetism, leading to the predominant use of electronic or nuclear spins in quantum magnetometers\,\cite{budker2007optical,degen2017quantum,mitchell2020colloquium}. Electron spins are leveraged to create the most sensitive magnetometers, while nuclear spins provide superior energy resolution due to their smaller gyromagnetic ratio\,\cite{vasilakis2009limits,graner2016reduced}. Spin-based magnetometers are at the cutting edge of precision magnetic field measurement technology, marking a significant advancement over classical magnetometers that rely on the magnetic properties of specific materials. Table \ref{tab:my_label1} details the specifications of various spin sensors, including spin species, sensitivity or precision, energy resolution, sensor size, and applications. Unlike other magnetometer technologies that may require cryogenic cooling or face significant size constraints, most spin-based magnetometers are compact and portable, operating effectively at room temperature. Their adaptability makes them invaluable in diverse fields such as geophysics, biomedicine, and fundamental physics. In the following, we will briefly review several key spin-based quantum sensors that have been employed in searching for exotic spin-dependent interactions.

\subsection{Optically pumped magnetometer} 

\begin{figure}[t]  
	\makeatletter
	\def\@captype{figure}
	\makeatother
	\centering
        \includegraphics[width=0.45\textwidth]{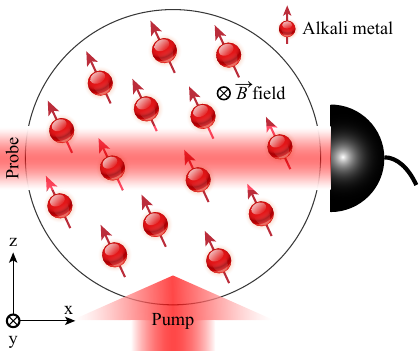}
	\caption{A basic schematic of an optically pumped magnetometer.
 Pump light polarizes the atoms, which allows the spins of these polarized atoms to evolve under the influence of an external magnetic field. The resultant state of the atoms’ polarization is detected by measuring transmission or polarization rotation of the probe light.}
 \label{SERFmag}
\end{figure}

Based on the discovery of optical pumping by Alfred Kastler, for which he was awarded the 1966 Nobel Prize in Physics\,\cite{kastler1950phys}, a prototype spin magnetometer was proposed that the strength of a magnetic field could be measured by detecting the precession of atomic spins in 1957\,\cite{dehmelt1957modulation}. That same year, the first Optical Pump Magnetometer (OPM) was realized\,\cite{bell1957optical}.
OPMs typically utilize ``optical pumping" to polarize alkali atoms such as potassium (K), rubidium (Rb) and cesium (Cs) within a vapor cell along a specific direction.
These spin-polarized atoms then precess at their Larmor frequency in an external magnetic field. 
The measurement of this precession frequency, along with changes in the optical properties of the atoms, such as absorption or polarization rotation, enables precise determination of the magnetic field strength, as depicted in Fig.\,\ref{SERFmag}.

The achievable sensitivity of OPMs is limited by the fundamental quantum-mechanical uncertainty in measuring atomic spin projections. The spin-projection-noise-limited (or atomic shot-noise-limited) sensitivity of a polarized atomic sample to magnetic fields is determined by  
\begin{equation}
\delta B=\frac{1}{\gamma_e \sqrt{n_e T_{2e} V t}},
\label{SERF}
\end{equation}
where $n_e$ is the density of atoms, $\gamma_e$ is their gyromagnetic ratio, $T_{2e}$ is the transverse relaxation time, and $V$ is the atomic volume interacting with light, and $t$ is the measurement time. This generally refers to the Standard Quantum Limit (SQL). Therefore, to achieve the highest possible precision in magnetic field measurements, it is beneficial to maximize the relaxation time of atomic polarization $T_{2e}$ as well as the largest possible number of atoms $N$.

Prior to the 2000s, research on resonant nonlinear magneto-optical effects was crucial in the field of OPM\,\cite{cohen1969detection,aleksandrov1987optically,budker2000sensitive,aleksandrov1995laser}.
A major breakthrough of OPM's sensitivity was achieved following the discovery of the Spin-Exchange Relaxation-Free (SERF) effect\,\cite{happer1973spin}. This effect occurs when the rapid spin-exchange collision rate surpasses the Larmor precession frequency of the atoms, thereby nearly eliminating spin-exchange relaxation, a typical process that limits sensitivity.  Consequently, this leads to markedly improved sensitivity.
The SERF magnetometers represent a new class of OPMs that operates at high enough atomic densities (exceeding $10^{14}$/cm$^{3}$ via heating) and low enough magnetic fields (less than 10\,nT). The first SERF magnetometer, realized in 2002, achieved an impressive sensitivity of 10\,fT/Hz$^{1/2}$\,\cite{allred2002high}. This sensitivity was further improved through the implementation of gradiometer configurations, with a notable achievement in 2003 where the sensitivity reached 0.54\,fT/Hz$^{1/2}$\,\cite{kominis2003subfemtotesla}.
In 2007, a SERF magnetometer using a MEMS (microelectromechanical systems) vapor cell achieved a sensitivity of 70\,fT/Hz$^{1/2}$\,\cite{shah2007subpicotesla}.
A significant milestone was reached in 2010 when the use of a ferrite magnetic shield enabled an ultra-high sensitivity of 0.16\,fT/Hz$^{1/2}$\,\cite{dang2010ultrahigh}, still remaining the record for gradient magnetic sensitivity at low fields. Current sensitivities of OPMs utilizing K, Rb and Cs are listed in Table\,\ref{tab:my_label1}. 
The projected fundamental limits of the SERF magnetometers based on spin projection noise are below $10^{-2}$ fT/Hz$^{1/2}$\,\cite{ji2018new,ji2023constraints}.

Due to their exploitation of electron spins, low frequency, and high sensitivity, the SERF magnetometers are primarily utilized to search for exotic electron spin-dependent interactions\,\cite{Afach:2021pfd}. 
These include axion-electron spin couplings for mass ranges below $10^{-12}$eV and exotic electron spin-dependent forces for force ranges greater than $10^{-2}$ meters, constrained by the size of the SERF magnetometer\,\cite{kim2019experimental,xiao2023femtotesla,xiao2024exotic,wu2022experimental,ji2018new,ji2023constraints}. Recently, a new method based on the calculation of atomic motion inside the vapor cell has enabled the setting of constraints for force ranges between $10^{-8}$ and $10^{-6}$ cm\,\cite{xiao2024exotic}.

\subsection{Atomic comagnetometer}
\label{atcom}

Utilizing electron spins, SERF OPMs have demonstrated the highest sensitivity at low frequency, 
while modern comagnetometers with the use of nuclear spins are the most sensitive sensors for measuring energy splitting between quantum states, leading to highly accurate measurements of anomalous magnetic-like fields\,\cite{vasilakis2009limits, wei2023ultrasensitive}. 
The initial application of nuclear spins in fundamental physics, which also marked the beginning of comagnetometry, occurred in 1960. Hughes\,\cite{hughes1960upper} and Drever\,\cite{drever1961search} analyzed magnetic resonance spectroscopy of the nucleus of lithium-7 ${ }^7 \mathrm{Li}$ and a proton, comparing them at various orientations of the  ${ }^7 \mathrm{Li}$ quadrupole in relation to the galactic center\,\cite{hughes1960upper,drever1961search}. Since then, comagnetometers have made significant progress in improving absolute energy sensitivity,
achieving energy sensitivities in the $10^{-26}$\,eV range for some applications\,\cite{vasilakis2009limits,graner2016reduced,terrano2021comagnetometer}.

\begin{figure}[t]  
	\makeatletter
	\def\@captype{figure}
	\makeatother
	\centering
        \includegraphics[width=0.45\textwidth]{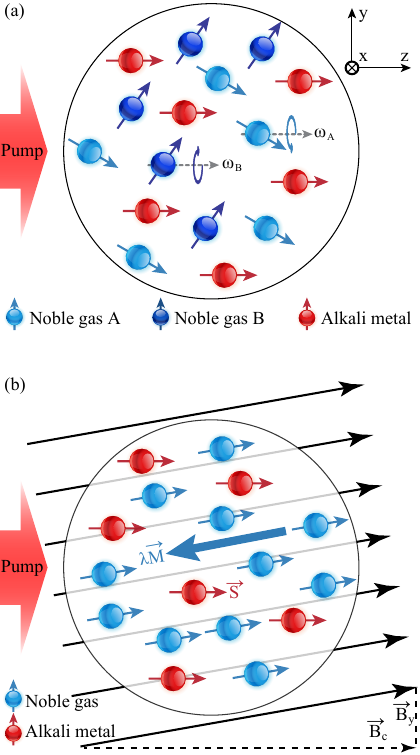}
	\caption{A basic schematic of atomic comagnetometers. (a) Clock-comparison comagnetometer that employs spatially-overlapped nuclear spins of two noble gases to measure the frequency shift caused by the pseudomagnetic fields. (b) Self-compensated comagnetometer that operates at a specific magnetic field $\vec{B}_c = - \vec{B}^{e} - \vec{B}^{n}$ (see text). When a disturbance magnetic field $B_y$ is applied, the nuclear magnetization $\vec{M}$ would adiabatically follow the slow changes in external magnetic field and thus cancel the external field disturbance.}
\label{Comag}
\end{figure}

Usually, spin-based comagnetometers simultaneously measure multiple species’ different responses based on their specific spin properties to effectively track the same external magnetic field.
Here we focus on nuclear-spin comagnetometers due to ultrahigh absolute energy sensitivity of nuclear spins\,\cite{wei2023ultrasensitive,vasilakis2009limits}. Generally, nuclear-spin comagnetometers also use optical pumping to generate a spin-polarized electronic state, then transferred to the nucleus\,\cite{walker1997spin,happer1973spin}. Typical comagnetometers involving two nuclear spin species include Mercury ${ }^{199} \mathrm{Hg}-{ }^{201} \mathrm{Hg}$ through the direct pumping and probing with UV light, noble gases ${ }^3 \mathrm{He}-{ }^{21} \mathrm{Ne}$, ${ }^3 \mathrm{He}-{ }^{129} \mathrm{Xe}$ and ${ }^{129} \mathrm{Xe}-{ }^{131} \mathrm{Xe}$  through collisional-exchange-polarization and optical-readout techniques\,\cite{walker1997spin,happer1973spin}. In most cases, the spin ensembles are spatially-overlapping, contained in a single chamber\,\cite{griffith2009improved,aleksandrov1983limitations,venema1992search,lamoreaux1987new,jacobs1993testing,peck2012limits,youdin1996limits,hunter2013using,hunter2014using,lee2018improved,vasilakis2009limits,almasi2020new,chupp1988precision,chupp1994spin}.
Noble gases are filled into alkali-metal vapor in the same cell where lasers pumped and probed the alkali-metal atoms, and then the nuclear spins in noble gases via spin-exchange collisions.
Some comagnetometers use spatially-separated spins where the spin ensembles are located in separate chambers\,\cite{allmendinger2014new,allmendinger2019measurement,gemmel2010limit}.
This allows the comparison of spins of the same species, or spin species which require significantly different environmental conditions.
In contrast to comagnetometers based on overlapping ensembles of multiple spin species, which typically encounter systematic errors stemming from magnetic field gradients\,\cite{sheng2014new},
comagnetometers based on the nuclear spins within an ensemble of identical molecules\,\cite{wu2018nuclear} or utilizing Zeeman transitions of the dual hyperfine levels in ground-state $^{87}$Rb atoms\,\cite{wang2020single} have been demonstrated, offering a novel approach to mitigate such issues.
According to comparison type, there are mainly two kinds of comagnetometers: clock comparison and quantization axis comparison.

Clock-comparison comagnetometers measure the procession frequencies $\omega_i$ of each spin ensemble $i$ and compare $\omega_i$ by a weighted frequency difference
\begin{equation}
\begin{array}{lll}
    \delta \omega(t) & = & \omega_A - \frac{\gamma_A}{\gamma_B} \omega_2  \\
    & = & (\gamma_A - \frac{\gamma_A}{\gamma_B} \gamma_B ) B(t) + (1 - \frac{\gamma_A}{\gamma_B} ) \omega_{\rm{n-mag}}(t) + ...
\end{array}
\end{equation}
which cancels out any dependence on magnetic field
fluctuations (the first term is zero)\,\cite{chupp1988precision,chupp1994spin,stoner1996demonstration,bear1998improved,rosenberry2001atomic,glenday2008limits,bulatowicz2013laboratory,gemmel2010limit,gemmel2010ultra,tullney2013constraints}.
Here $\gamma_{A/B}$ are their gyromagnetic ratios, $B(t)$ is the magnetic field, and $\omega_{\rm{n-mag}}(t)$ arises from the precession induced by non-magnetic fields, such as those predicted by new physics beyond the SM, e.g.,  $\vec{\mathcal{O}}$ in
Eq.\,\eqref{hamD}, or inertial rotation.
Figure\,\ref{Comag}(a) illustrates the basic operational principle of the clock-comparison comagnetometer.
The sensitivity of the measurement to pseudomagnetic fields, derived from the frequency shift, can be quantified as\,\cite{ledbetter2005progress} 
\begin{equation}
     \delta \omega =\frac{1}{\sqrt{{n_n} V T_{2n} t}},
\end{equation}
where $n_n$ is the density of nuclear-spin species, $T_{2n}$ is the transverse relaxation time of nuclear spins, and $V$ is the atomic volume interacting with light, and $t$ is the measurement time.
Typical comagnetometers using $^{199}$Hg-$^{133}$Cs achieved a frequency precision of $\delta \omega \approx 0.8\,\mu$Hz\,\cite{peck2012limits},
while those using $^{199}$Hg-$^{201}$Hg attained a frequency precision of $\delta \omega \approx 0.5\,\mu$Hz\,\cite{venema1992search}.


Another type of comagnetometers is the self-compensating ones that compare the spin-quantization axes of co-located spin ensembles. These devices typically involve one spin ensemble comprising an alkali-metal vapor and another consisting of a noble gas, specifically K-$^3$He and Rb-$^{21}$Ne\,\cite{kornack2002dynamics,kornack2005nuclear,smiciklas2011new,vasilakis2009limits,lee2018improved,almasi2020new,wei2023ultrasensitive}.
Figure\,\ref{Comag}(b) illustrates the basic operating principle of this type of comagnetometer. In these systems, the polarized electron spins of the alkali-metal atoms are coupled with the nuclear spins of the noble gases through spin-exchange interactions. This coupling generates an effective magnetic field $\vec{B}^{e/n} = \lambda \vec{M}^{e/n}  = (8 \pi \kappa_0 /3) \vec{M}^{e/n} $ where $\vec{M}^{e/n}$ represents the electron or nuclear magnetizations and the enhancement factor $\kappa_0$ for these interactions varies from 5 to 600, depending on the alkali-metal-noble-gas pairs\,\cite{walker1997spin,happer1973spin}. When a bias magnetic field $\vec{B}_c$ applied along the pump direction is adjusted to the self-compensating point, i.e., $\vec{B}_c = - \vec{B}^{e} - \vec{B}^{n}$, 
the nuclear magnetization $\vec{M}^{n}$ adiabatically tracks a slowly changing normal magnetic field.
This configuration renders the alkali magnetometer insensitive to normal magnetic fields transverse to the pumping direction, while its sensitivity to anomalous fields, such as those arising from non-magnetic phenomena predicted by theories beyond the SM, is preserved\,\cite{kornack2002dynamics,kornack2005nuclear,smiciklas2011new,vasilakis2009limits,lee2018improved,almasi2020new,wei2023ultrasensitive,terrano2021comagnetometer}. 
This cancellation is only effective at frequencies below the Larmor frequency of the nuclear spins.
This also allows high-sensitivity SERF operation of the OPM since the net magnetic field experienced by the alkali-metal atoms is close to zero\,\cite{kornack2005nuclear}.
In this regime, the sensitivity to pseudomagnetic fields is the same as that defined by Eq.\,\eqref{SERF}\,\cite{kornack2002dynamics,kornack2005nuclear,smiciklas2011new,vasilakis2009limits,lee2018improved,almasi2020new,wei2023ultrasensitive}
where the nuclear spins share the same magnetic sensitivity with the electron spins.
Here we would like to emphasize that the self-compensating comagnetometer exhibits insensitivity to normal magnetic fields, typically at frequencies below 1\,Hz, with Eq.\,\eqref{SERF} pertaining specifically to pseudo-magnetic fields.
For instance,
the K-$^3$He comagnetometer achieved a pseudo-magentic field sensitivity of 0.75\,fT/Hz$^{1/2}$ at around 0.18\,Hz\,\cite{vasilakis2009limits}.
A recent study has unveiled a new type of atomic comagnetometer based on a self-compensation mechanism originating from the destructive interference between alkali-metal and noble-gas spins.
Remarkably, this new comagnetometer employing K-$^3$He system has achieved a significant reduction in magnetic noise at higher frequencies up to 100\,Hz.


The primary objective behind comagnetometers is to mitigate common-mode noise that affects magnetic field measurements. Consequently, their advancement has been fueled by a growing interest in fundamental physics experiments, including tests of the equivalence principle and searches for a permanent EDM of particles, which could potentially unveil new physics beyond the SM\,\cite{abel2020measurement,allmendinger2019measurement}. 
 Various comagnetometer implementations have been developed for fundamental physics applications, with a recent comprehensive review focusing on their utility in probing dark matter and new physics\,\cite{terrano2021comagnetometer}. Furthermore, comagnetometers are explored as a component of inertial navigation systems (gyroscopes) in scenarios where the global positioning system (GPS) is unreliable or inaccessible, such as in submarine navigation and space exploration.

\subsection{Spin maser}

Masers generate coherent electromagnetic waves (microwaves) through amplification by stimulated emission.
Since the first maser was built in 1953\,\cite{gordon1955maser}, masers have been successfully implemented across various systems.
Notably, spins play a pivotal role in the operation of masers\,\cite{bear1998improved,bear2000limit,chupp1988precision,inoue2011experimental,inoue2011frequency,sato2018development,stoner1996demonstration,oxborrow2012room,breeze2018continuous,kraus2014room}.
The frequency resolution of maser is unaffected by spin decoherence but is instead constrained by the maser's stability.
Thanks to their exceptional frequency stability, spin masers facilitate highly accurate measurements of frequency variations induced by interactions with external electromagnetic fields. The nuclear spin maser, utilizing $^3$He and optical pumping for the metastable state of $^3$He, was first achieved in 1964\,\cite{gentile2017optically}. 

\begin{figure}[t]  
	\makeatletter
	\def\@captype{figure}
	\makeatother
	\centering
        \includegraphics[width=0.45\textwidth]{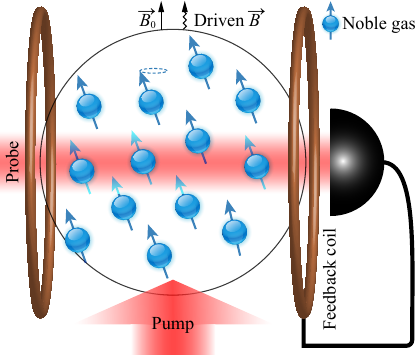}
	\caption{A basic schematic of noble-gas spin maser\,\cite{sato2018development}. Noble-gas nuclear spins are polarized and detected by spin exchange with optically pumped alkali-metal spins. In the presence of a bias field $B_0$, these polarized noble-gas spins are coupled to a feedback circuit, which provides real-time transverse magnetic field feedback. Once the maser threshold conditions are met, a stable nuclear spin maser is achieved. For a Floquet maser\,\cite{jiang2021floquet}, an oscillating magnetic field, aligned with the bias field, is introduced, enabling the observation of a multimode maser oscillating at frequencies corresponding to transitions between Floquet states.}
 \label{maser}
\end{figure}

Figure\,\ref{maser} presents a basic schematic of a noble-gas spin maser. This device operates based on the principle of spin-exchange pumping, in which population inversion is generated through spin exchange interactions between the noble gas and optically polarized alkali-metal atoms\,\cite{sato2018development,bear1998improved,bear2000limit,stoner1996demonstration,chupp1988precision,chupp1994spin}.
The magnetization of nuclear spins is detected in-situ by an alkali-metal magnetometer, which provide positive feedback to the noble-gas transitions.
Similar to atomic comagnetometer discussed in Sec.\,\ref{atcom},
a dual-species noble-gas maser (e.g., $^{3}$He -$^{129}$Xe was further
introduced in 1994\,\cite{chupp1994spin},
effectively reducing common-mode systematic effects, such as fluctuations in the uniform magnetic field.
This dual-species strategy facilitates simultaneous active oscillations across different noble-gas species, allowing for sensitive differential measurements of Zeeman transition frequencies, thereby enhancing the maser's precision and stability.
For example, the dual-species maser using $^{129}$Xe-$^{131}$Xe has achieved frequency precision up to 6.2\,$\mu$Hz,
as detailed in Table\,\ref{tab:my_label1}.

While the dual-species maser significantly mitigates common-mode magnetic noise,
it faces challenges in reducing additional magnetic effects, namely, the non-common magnetic effects generated by in-situ alkali-metal spins on different types of noble-gas isotopes. As a result, the differential frequency resolution of a single-chamber dual-species maser is considerably constrained by the effective magnetic field produced by optically pumped alkali metals. To overcome this limitation, some designs adopt a dual-chamber configuration, which separates the spin-exchange pumping process from the maser operations\,\cite{chupp1988precision,chupp1994spin,stoner1996demonstration,bear1998improved}. In such setups, the noble-gas spins in the second chamber are typically detected using external magnetometers, including pickup coils and SQUID magnetometers.
For instance, as shown in Table\,\ref{tab:my_label1}, the dual-chamber dual-species maser involving $^{129}$Xe-$^3$He has achieved frequency precision of 6.1\,nHz.

A recent advancement introduced a novel maser variant, namely, the Floquet maser, which leverages a sequence of time-independent Floquet states and energy levels\,\cite{jiang2021floquet}. 
This maser operates under an additional magnetic field applied along the $z$ axis, periodically driving the noble-gas spin system and establishing a Floquet system characterized by time-dependent Hamiltonians.
This system features a set of time-independent Floquet states and energy levels, akin to the artificial dimension found in the Brillouin zone\,\cite{shirley1965solution}. Beyond transitions among inherent states, transitions between Floquet states appear as sidebands, which can be tailored by varying the frequency and amplitude of the periodic drive.
Distinct from traditional frequency shift measurements, the Floquet maser assesses the amplitude of Floquet transitions with a sensitivity reaching 700\,fT/Hz$^{1/2}$ below 60\,mHz.


Because the frequency of radiowaves produced by masers is highly stable, these devices enable exquisitely sensitive measurement of their frequency shifts caused by the interactions with external electromagnetic fields.
This opens up exciting possibilities for developing precise metrology in fundamental physics, including
Lorentz symmetry and charge-parity-time
violation\,\cite{bear2000limit,bear2000fundamental},
EDMs\,\cite{rosenberry2001atomic,sato2018development,inoue2011experimental,inoue2011frequency},
spin-dependent forces between neutrons using $^3$He-$^{129}$Xe maser\,\cite{glenday2008limits},
and ultralight dark matter\,\cite{jiang2021floquet}.

\subsection{Nuclear magnetic resonance}
\label{NMR-amp}


Nuclear magnetic resonance (NMR) was initially described and measured in molecular beams by Rabi in 1938, which earned him the Nobel Prize in Physics in 1944\,\cite{rabi1938new}. 
Subsequent advancements by Ramsey, Bloch, Purcell, Ernst, Lauterbur and Mansfield further developed NMR techniques, leading to their collective Nobel Prize accolades\,\cite{alvarez1940quantitative,bloch1946nuclear,ramsey1950molecular,aue1976two,lauterbur1973image,mansfield1977multi}. 
Leveraging the principles of quantum-mechanical resonance, NMR facilitates highly precise detection of tiny energy shifts induced by exotic spin-dependent interactions. 
Various experiments have been proposed or demonstrated to investigate such exotic spin-dependent interactions using NMR, notably the Cosmic Axion Spin Precession Experiment (CASPEr)\,\cite{budker2014proposal,garcon2017cosmic,kimball2020overview, aybas2021search, garcon2019constraints, wu2019search} and the Axion Resonant InterAction Detection Experiment (ARIADNE)\,\cite{aggarwal2022characterization,arvanitaki2014resonantly}.

\begin{figure}[t]  
	\makeatletter
	\def\@captype{figure}
	\makeatother
        \centering
        \includegraphics[width=0.45\textwidth]{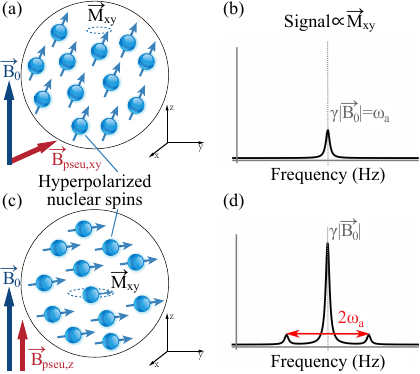}
	\caption{The principle of nuclear magnetic resonance (NMR)-based searches. (a) and (b), NMR resonant scheme. Hyperpolarized nuclear spins are situated within a bias field $\vec{B}_0$, which is finely tuned to match the frequency $\omega_a$ of an external pseudomagnetic field $\vec{B}_{pseu,xy}$ generated either by ALP dark matter (see CASPEr project\,\cite{budker2014proposal,garcon2017cosmic,kimball2020overview}) or by exotic spin-dependent forces (see ARIADNE project\,\cite{arvanitaki2014resonantly}). When the spin Larmor frequency is precisely matched to the frequency of the pseudomagnetic field, i.e., $\gamma |B_0|=\omega_a$, it results in the induction of transverse nuclear magnetization  $\vec{M}_{xy}$, which can be measured using an external detector, such as a SQUID magnetometer. (c) and (d), NMR sideband scheme. In this configuration, the pseudomagnetic field is aligned parallel to the bias field. When the pseudomagnetic field's frequency is significantly below the spin Larmor frequency, NMR sidenbands can be generated. The figure is reproduced from Ref.\,\cite{garcon2017cosmic}.}
 \label{Figure6}
\end{figure}


{A pick-up loop, a SQUID or an atomic magnetometer, can be used to measure the dipole-dipole interactions generated by a large number of polarized nuclear spins (see Fig.\,\ref{Figure6}).
The magnetic sensitivity decreases according to an inverse cubic relationship, expressed as $1/r^{3}$, where $r$ denotes the distance between nuclear sample and the detector. Assume that the external magnetometer is on the surface of the nuclear sample, the sensitivity of
a polarized nuclear sample to magnetic fields is determined by\,\cite{budker2014proposal,garcon2017cosmic}
\begin{equation}
    \delta B \approx \frac{ \delta B_{\textrm{d}}}{\eta \sqrt{t}}, \,\,\, \eta = \frac{8 \pi}{3} n_n \mu_n P_0^n \gamma_i  T_{2n},
    \label{Bnmr}
\end{equation}
where $\delta B_{\textrm{d}}$ denotes the magnetic sensitivity of the external NMR detector (i.e., SQUID magnetometer),
$\eta$ refers to the magnetic amplification factor, $n_n$ is the number density of nuclear spins,
$\mu_n$ denotes the magnetic moment of the nucleus,
$P_0^n$ is the equilibrium polarization of nuclear spins, $\gamma_i$ is the gyromagnetic ratio of the nucleus,
and $T_{2n}$ denotes the coherence time of nuclear spins.
The amplification factor $\eta$ has the potential to be substantial.
For example, in the fully polarized liquid ${ }^{129} \mathrm{Xe}$ ($P_0^n=1$), with a spin density of nearly $n_n \approx 10^{22}$ spins$/ \mathrm{cm}^3$ and $T_{2n} \gtrsim 1000 \mathrm{~s}$,
the resulting amplification factor is impressive $\eta \gtrsim 10^6$.
Utilizing a SQUID magnetometer as the external NMR detector with a detection sensitivity of about $\delta B_{\rm{d}} \approx 1 \mathrm{~fT} / \mathrm{Hz}^{1/2}$,
the measurement sensitivity for the input magnetic field can be enhanced to $10^{-6} \mathrm{~fT} / \mathrm{Hz}^{1/2}$.

CASPEr experiments employ either liquid or solid states of nuclear spins, such as liquid $^{129}$Xe or solid $^{207}$Pb, to increase the spin density\,\cite{aybas2021search,garcon2017cosmic,wu2019search}.
The noble-gas nuclear spins ($^{129}$Xe and $^{3}$He) are typically polarized through the spin-exchange optical pumping with optically polarized alkali metal atoms in gas phase\,\cite{happer1973spin,walker1997spin}.
In CASPEr, the hyperpolarized $^{129}$Xe is first condensed into solid form inside a region cooled by a liquid-nitrogen bath in the presence of a leading magnetic field, then sublimated to gas form, and finally compressed into a liquid state by a piston. 
Solid-state $^{207}$Pb spins in a ferroelectric PMN-PT crystal [$(\mathrm{PbMg}_{1 / 3} \mathrm{Nb}_{2 / 3} \mathrm{O}_3)_{2 / 3}-(\mathrm{PbTiO}_3)_{1 / 3}$] are polarized by applying a voltage across its faces at room temperature\,\cite{aybas2021search}.
By scanning the bias field, resonance occurs when the Larmor frequency of nuclear spins matches the oscillating frequency of an external field.
In this case, the spins are tilted away from the direction of the bias field, precess, generate a time-dependent magnetization that can be measured\,\cite{budker2014proposal,walker1997spin,garcon2017cosmic}. 
By adjusting the bias field,
CASPEr can potentially cover the frequency range of $10^{3}$-$10^{8}$\,Hz\,\cite{budker2014proposal,kimball2020overview}.
The CASPEr experiments are currently in progress, with efforts being made to prepare high-polarization liquid $^{129}$Xe sample and prolong its coherence time.

ARIADNE highlights a novel method for detecting axion-mediated forces, combining techniques from NMR and short-distance gravity tests\,\cite{arvanitaki2014resonantly, aggarwal2022characterization}. The method is sensitive to Peccei-Quinn (PQ) axion decay constants between $10^9$ and $10^{12}$\,GeV or axion masses between $10^{-6}$ and $10^{-3}$\,eV. 
The experiment leverages the resonant coupling between an unpolarized source mass's rotational frequency and a gas of hyperpolarized $^{3}$He nuclear spins' matching spin precession frequency, measured by a SQUID. 
The projected constraint presents a significant improvement over current experimental limits, potentially being up to 8 orders of magnitude more sensitive, and bridges the gap between astrophysical bounds and cosmic PQ axion searches\,\cite{raffelt2008astrophysical}.
The ARIADNE experiment is currently in progress, with efforts being made to analysis systematic errors\,\cite{aggarwal2022characterization}.



Apart from conventional NMR techniques, a new possibility at ultralow magnetic fields has recently emerged, based on zero- to ultralow-field (ZULF)-NMR\,\cite{wu2019search,wu2022experimental,garcon2017cosmic,garcon2019constraints}. 
Compared to the resonant detection, ZULF experiments are suitable to measure non-resonant magnetic fields or pseudomagnetic fields.
As illustrated in Fig.\,\ref{Figure6}(c) and (d), a pseudomagnetic field $\vec{B}_{\rm{pseu,z}}$ aligned with the bias field periodically modulates the Larmor frequency of nuclear spin.
This modulation gives rise to two sidebands flanking the central Larmor frequency, separated by a modulation frequency $\omega_a$\,\cite{garcon2017cosmic,garcon2019constraints}.
In current CASPEr-ZULF experiments\,\cite{garcon2019constraints},
the sensitivity is mainly limited by the nuclear-spin polarization of samples.
In the near future, by using parahydrogen-induced polarization (PHIP) technique,
the certain nuclear spin polarization can be greatly improved, for example,
${ }^{15} \mathrm{N},{ }^{13} \mathrm{C}_2$-acetonitrile $\left({ }^{13} \mathrm{CH}_3^{13} \mathrm{C}^{15} \mathrm{N}\right)$ may potentially achieve nuclear spin polarizations on the order of $P_0^n\approx 0.1$\,\cite{theis2011parahydrogen}.


The combination of hyperpolarization techniques with NMR would significantly enhance the sensitivity of nuclear spins to magnetic fields.
This provides a powerful technique for probing exotic spin-dependent interactions.
There is an increasing interest within the scientific community in harnessing NMR techniques to probe into the realms of ALP dark matter\,\cite{budker2014proposal,garcon2019constraints,wu2019search,aybas2021search} and spin-dependent forces\,\cite{arvanitaki2014resonantly}.
For example,
the CASPEr-ZULF NMR has explored ALP dark matter with masses ranging
from $10^{-22}$ to $1.3\times 10^{-17}$\,eV\,\cite{wu2019search}, and with masses ranging from
$1.0 \times 10^{-16}$ to $7.8 \times 10^{-14}$\,eV\,\cite{garcon2019constraints}.
Additionally, through the precision measurement
of $^{207}$Pb solid-state NMR,
new constraints have been established for ALP-nucleon and ALP-gluon interactions across a mass window of $1.62-1.66 \times 10^{-7}$ eV\,\cite{aybas2021search}.
The ARIADNE project proposes a NMR-based experiment to search for spin-dependent forces mediated by axions within the mass range of $10^{-6}$\,eV and $10^{-3}$\,eV.}



\subsection{Spin gas amplifier}


As introduced in Sec.\,\ref{NMR-amp},
hyperpolarized NMR holds the potential for significant magnetic field amplification.
However, efforts to demonstrate this type of amplification are ongoing.
Such works have explored scenarios where nuclear spins are measured in close proximity to atomic and SQUID magnetometers. In such setups, achieving high spin polarization in nuclear spins and maintaining sensitive readout proves experimentally challenging.
In contrast to these approaches, an alternative method has been developed where nuclear spins and the detector coexist within the same vapor cell, achieving considerable magnetic amplification ranging from 100 to 5400\,\cite{su2021search,jiang2021search,jiang2022floquet,wang2022limits,wang2023search,jiangdark2024}. This method capitalizes on the substantial Fermi-contact enhancement factor to boost nuclear spin signals, which are detected in situ by an atomic magnetometer.
For a heavy noble gas, such as $^{129}$Xe, the magnetic field generated by nuclear magnetization can be enhanced by a large factor of 540\,\cite{walker1997spin}.
Consequently, this approach employs hyperpolarized, long-lived nuclear spins as a pre-amplifier, effectively boosting an external resonant oscillating field, as demonstrated by recent studies\,\cite{su2021search,jiang2021search,jiang2022floquet,wang2022limits,wang2023search,jiangdark2024}.

\begin{figure}[t]  
	\makeatletter
	\def\@captype{figure}
	\makeatother
        \centering
        \includegraphics[width=0.45\textwidth]{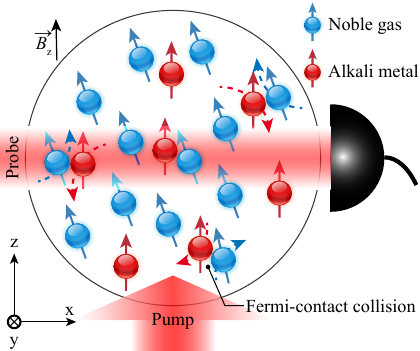}
	\caption{The principle of spin amplifier. The bias field is tuned to match the frequency of external magnetic field.}
\label{Spinamp}
\end{figure}

The basic principle of this type of spin gas amplifier is outlined below.
The resonant measured fields collectively excite ensembles of polarized $^{129}$Xe-sensor spins, corresponding classically
to a tilt of the collective spin about the polarization axis as shown in Fig.\,\ref{Spinamp}. 
Based on the Fermi-contact interactions, the effective field generated by the transverse nuclear magnetization is measured by the in-situ alkali-metal magnetometer.
The shot-noise-limited sensitivity of spin amplifier is\,\cite{jiang2021search}
\begin{equation}
	 \delta B=\frac{1}{\eta \gamma_e \sqrt{n_e T_{2e} V t}},\,\,\, \eta\approx \dfrac{8\pi}{3}\kappa_0 n_n \mu_n P^{n}_{0} \gamma_{i} T_{\textrm{2n}},
	 \label{beff}
\end{equation}
where $\kappa_0$ is the Fermi-contact enhancement factor\,\cite{walker1997spin},
$n_n$ is the density of nuclear spins, $\mu_n$ is the neutron magnetic moment, $P^n_0$ is the equilibrium polarization of noble-gas nucleus,
$\gamma_i$ is the gyromagnetic ratio of nucleus,
and $T_{2n}$ is the nuclear-spin coherence time.
The factor $\eta$ is the amplification gain of the spin amplifier to external magnetic fields.
By comparing Eq.\,\eqref{beff} with Eq.\,\eqref{Bnmr}, we identify two distinct differences between the spin gas amplifier and previous NMR amplification methods.
First, the in-situ detection of noble-gas spins leads to an enhancement of the amplification factor by a multiple of $\kappa_0$;
secondly, the spin gas amplifier utilizes an in-situ OPM for noble-gas spin detection, which relies on Eq.\,\eqref{SERF} as its shot-limited sensitivity, in contrast to external detectors like SQUID magnetometers.


It is essential to acknowledge that various parameters, including $\{\kappa_0, n_{n}, P^{n}_{0}, T_{2n} \}$,
play crucial roles in attaining a high amplification factor.
In a demonstrated experiment, the coherence time of $^{129}$Xe nuclear spins was about $T_{2n}\approx 20$\,s,
the polarization of $^{129}$Xe reached approximately $P_0^n \approx 0.3$,
and an amplification factor of $\eta \approx 128$ was achieved\,\cite{jiang2021search},
leading to a prototype spin amplifier's sensitivity of 18\,fT/Hz$^{1/2}$.
Subsequently, advancements in the dark spin technique have further elevated the amplification factor for $^{129}$Xe to $\eta \approx 5400$\,\cite{jiangdark2024},
enhancing the sensitivity to 3\,fT/Hz$^{1/2}$.
The potential for further improvement in the amplification factor $\eta$ remains substantial.
For example,
in the $^3$He-K system,
a fully polarized sample of $^{3} \mathrm{He}$ gas with $n_n \approx 10^{20}$ spins$/\mathrm{cm}^3$ and $T_{2n}\gtrsim 1000\,\mathrm{s}$ can achieve an amplification factor $\eta \gtrsim 10^6$.
Moreover, the K OPM has demonstrated a high sensitivity of 10\,fT/Hz$^{1/2}$ in $^3$He-K system\,\cite{kominis2003subfemtotesla, kornack2005nuclear}.
As a result,
the projected sensitivity of $^3$He-K can achieve $10^{-5}$\,fT/Hz$^{1/2}$, which exceeds the fundamental sensitivity of SERF magnetometers and comagnetometers.
Assuming a shot-noise limited sensitivity of $0.01$\,fT/Hz$^{1/2}$ for K OPM,
the fundamental sensitivity could reach as high as $10^{-8}$\,fT/Hz$^{1/2}$.

While significant amplification can boost the response signal,
it is critical to consider the noise introduced by the atomic amplification process itself.
Generally, the sources of output noise include environmental noise, the amplifier medium, and detector-related noise. Specifically, the noise from the amplifier medium consists of the spin projection noise of nuclear spins.
Detector noise encompasses photon shot noise, spin projection noise of alkali-metal atoms, and electronic noise.
These types of noise can be less substantial than the magnetic noise originating from the magnetic shield, which is typically on the order of fT/Hz$^{1/2}$.
The primary challenge in achieving ultrahigh sensitivity with a spin amplifier lies in suppressing the magnetic field noise from the shields, given that nuclear spins can also amplify magnetic noise.

Analogous to the Floquet maser\,\cite{jiang2021floquet},
amplification can be achieved between Floquet states and energy levels\,\cite{jiang2022floquet}.
By harnessing synthetic dimensions facilitated by Floquet states,
the external magnetic field can undergo amplification at a sequence of comb-like frequencies, corresponding to multiple photon transitions.
Notably, the amplification effect was observed concurrently at multiple Floquet transitions, even though the frequency of the external fields was aligned with only one Floquet transition. Implementing Floquet amplification for magnetic field sensing, capable of concurrently measuring fields at various Floquet transitions, results in an improvement by an order of magnitude in the detection bandwidth achievable by spin amplifiers.

A novel technique leveraging a quantum Spin Amplifier for Particle Physics REsearch (SAPPHIRE)\,\cite{wang2022limits,wang2023search,su2021search} has been introduced for probing exotic nuclear spin-dependent interactions, including those involving axion dark matter and exotic spin-dependent forces. Through the adjustment of the bias field, this spin amplifier is adept at detecting signals within the 1-1000\,Hz range, making it particularly effective for investigating the high-frequency parameters of exotic spin-dependent interactions. The spin amplifier is especially well-suited for the exploration of exotic velocity-dependent interactions,
where the intensity of the pseudomagnetic field, induced by such interactions, is directly proportional to velocity.
Recent review has delved into the spin amplifier and its broad spectrum of applications\,\cite{su2022review}.


\subsection{Nitrogen-vacancy center}
The Nitrogen-vacancy(NV) center is a point defect in diamond consist of a substitutional nitrogen atom with an adjacent vacancy. The magnetic resonance studies on ensemble of NV centers in diamonds date back to the 1960s\,\cite{du1965electron}. In 1997, researchers successfully observed the fluorescence from a single NV center in diamond and captured its magnetic resonance spectrum\,\cite{Gruber1997}. 
In 2008, NV centers were recognized as potential magnetic field sensors, which were subsequently demonstrated both in single NV center and ensemble of NV centers experimentally\,\cite{Taylor2008b}. Since then, NV center in diamond has attracted lots of interest and have been applied in various applications\,\cite{Schirhagl2014,Du2024single}.

\begin{figure}[t]
    \centering
    \includegraphics[width=1\linewidth]{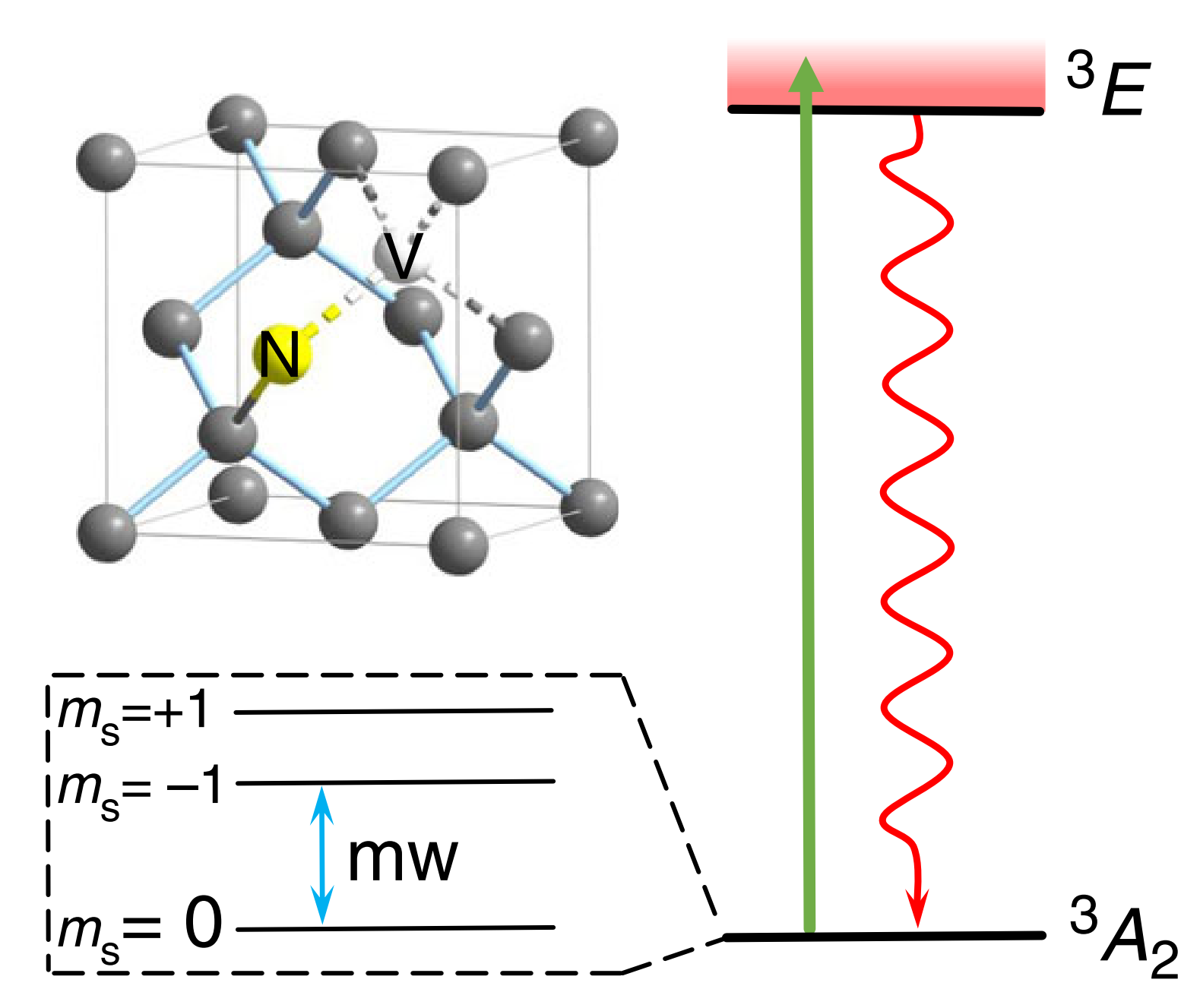}
    \caption{The principle of optical magnetic resonance of nitrogen-vacancy (NV) center in diamond. The figure is from Ref.\,\cite{rong2018searching}.}
    \label{fig:NVcenterenergylevel}
\end{figure}

The principle of optically detected magnetic resonance, atomic structure, and energy levels of the NV center in diamond are illustrated in Fig.\,\ref{fig:NVcenterenergylevel}. The zero-phonon line of NV center ground state ($^3$A$_2$) and excited states ($^3$E) is 637 nm. The NV center can be excited from the ground state to excited states by laser pulse, and decays back to ground state emitting photoluminescence. There also exists a nonradiative decay through intersystem crossing mechanism, resulting in optical spin initialization into the $|m_S=0\rangle$ ground state. In addition, the spin state of the NV center can be readout by detecting the spin-state-dependent fluorescence. Typically, a 532 nm laser is utilized to polarize the NV center, and photoluminescence around 630 nm is collected for spin state readout. This technique, known as optically detected magnetic resonance, enables the effective optical initialization and readout of the NV electron spins. 
The spin states $|m_S=0\rangle$ and $|m_S=1\rangle$ of ground state can be encoded as quantum sensor, with resonant microwaves being used to manipulate the quantum sensor. With the advance in quantum control of the single NV center, NV centers in diamonds have attracted wide attention and research in the fields of quantum information processing\,\cite{Childress2013} and quantum sensing\,\cite{taylor2008high,Schirhagl2014,Du2024single}.

NV centers in diamonds are capable of functioning in complex environments without requiring specific conditions such as cryogenics and vacuum systems. The atomic size of the single NV sensor enables it to achieve nanometer-scale spatial resolution. By increasing the number of sensing spins for statistical averaging, the ensemble-NV-magnetometer benefits from improved sensitivity. Currently, the narrowband AC sensitivity of ensemble-NV-magnetometers can reach 0.21\,pT/Hz$^{1/2}$\,\cite{barry2023sensitive}.

The shot noise limited sensitivity of NV ensemble magnetometer employing continuous-wave method can be estimated by\,\cite{Barry2017}
\begin{equation}
 \delta B=\frac{4}{3 \sqrt{3}} \frac{h}{g_e \mu_B} \frac{\Delta \Gamma }{C_{\mathrm{C W}} \sqrt{Rt}},
\end{equation}
where $R$ denotes photon-detection rate, $\Delta \Gamma$ is linewidth of CW spectrum, $C_{\mathrm{C W}}$ is contrast of CW spectrum, $h$ is the Planck constant, $g_e$ is the electronic g factor of NV center,  $\mu_B$ is the Bohr magneton.
As typical examples,
the single NV sensor achieved a sensitivity of 0.5\,nT/Hz$^{1/2}$\,\cite{Zhao2023} and the ensemble NV sensor demonstrated sensitivities of 0.21\,pT/Hz$^{1/2}$ for AC fields and 0.46\,pT/Hz$^{1/2}$ for DC fields\,\cite{barry2023sensitive}.

Due to its spatial resolution of the NV center, the NV center is well-suited for detecting exotic spin-dependent interactions in the micrometer scale.
Single NV centers in diamond have emerged as a solid-state spin quantum sensor to search for exotic spin-dependent interactions at the micrometer scale, exploiting the ability to enable close proximity between the sensor and the source. 

\subsection{Spin interferometry}

\begin{figure}[b]  
	\makeatletter
	\def\@captype{figure}
	\makeatother
        \centering
        \includegraphics[width=0.49\textwidth]{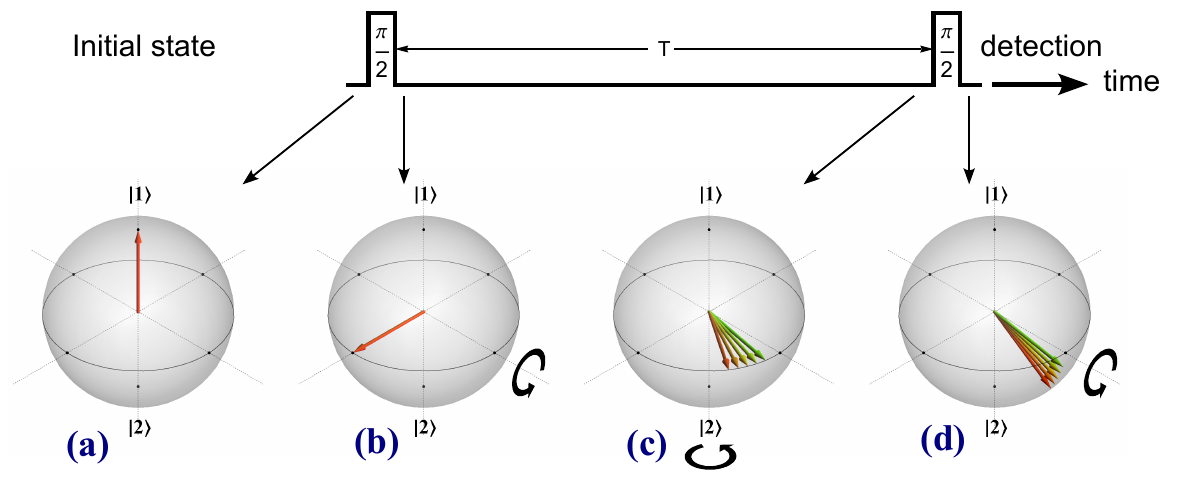}
	\caption{Ramsey-type spin interferometry. The figure is adapted from Ref.\,\cite{ivannikov2017ultracold}.}
 \label{ramsey}
\end{figure}

Spin interferometry, a sophisticated technique that exploits the quantum mechanical property of spin, has evolved significantly since its conceptual inception in quantum physics. The interferometry has roots dating back to the early 19th century, with the famous Young's double-slit experiment in 1801, marked the beginning of optical interferometry. As technology and theoretical understanding progressed, the principles of interferometry were expanded to the fundamental constituents of matter, such as atoms or intrinsic spin of particles, leading to the development of atomic interferometry\,\cite{cronin2009optics} or spin interferometry\,\cite{ramsey1949new,ramsey1950molecular} in the late 20th century. 

Spin interferometry involves the manipulation and measurement of the quantum spin states of particles like electrons or nuclei to detect phase shifts caused by external influences, allowing for the detection of extremely subtle effects such as minute magnetic fields or exotic forces predicted by theories beyond the SM. The basic principle hinges on the quantum mechanical phenomenon of superposition, where a particle's spin state can be in multiple states simultaneously. As shown in Fig.\,\ref{ramsey},
Ramsey-type spin interferometer splits a coherent spin state into a superposition of two different states $m_{S,1}$ and $ m_{S,2}$ by the first $\pi/2$ pulse, which then are allowed to evolve freely for a period of time $\tau$. During this interval, the two components of the superposition accumulate a phase difference due to their energy difference: 
\begin{equation}
     \phi = (E_{m_{S,1}} - E_{m_{S,2}}) \tau/\hbar.
\end{equation}
Following the free evolution period, a second $\pi/2$ pulse is applied to mix the states again, converting the phase difference accumulated during the free evolution into measurable population differences $N_{m_{S,1}}-N_{m_{S,2}}$ between the two spin states\,\cite{piegsa2009quantitative,ramsey1986neutron,ramsey1990experiments}. The energy difference can be induced by a variety of interactions, including magnetic fields, gravitational fields, or even theoretical interactions predicted by extensions to the SM of particle physics.

The sensitivity of spin interferometry lies in its ability to detect incredibly minute changes in these phases, which can be indicative of very faint forces or fields. The standard quantum limit for the phase shift measurement is $\delta \phi = \frac{1}{\hbar\sqrt{N t}}$ as well as the limit of the energy difference  measurement  $\delta E = \frac{\hbar}{ \tau \sqrt{N t}}$. Here $N$ is the total number of uncorrelated polarized particles and $t$ is the measurement time. The precision measurement of the energy difference means extremely precise determinations of small frequency changes. This principle is exploited in atomic clocks, where the spin interferometer is used to lock the frequency of a microwave oscillator to the frequency of a hyperfine transition in cesium or rubidium, forming the basis of the international definition of the second.
Apart from atomic clocks, Ramsey-type interferometry has been widely used in fundamental tests of physics, e.g., searches for EDMs\,\cite{piegsa2014ramsey,ramsey1986neutron}.
As typical examples,
the use of $\mathrm{HfF}^{+}$ molecular ions attained a frequency precision of 22.8\,$\mu$Hz\,\cite{roussy2023improved},
and the use of ThO reached a frequency precision of 59.4\,$\mu$Hz\,\cite{acme2018improved}.


\subsection{Other spin sensors}

In addition to the spin sensors mentioned earlier, a diverse range of other spin sensors are currently used to detecting exotic spin-dependent interactions, such as trapped ions.
Trapped ions were used to measure the frequency shift of Zeeman sublevel interacting with these interactions\,\cite{wineland1991search}, as well as potential spin-dependent forces between two trapped ions\,\cite{kimball2017constraints}.
Numerous avenues exist for further improving the fundamental and practical sensitivity of spin-based magnetometers.
Recently, a single-domain spinor BEC of ${ }^{87} \mathrm{Rb}$ was demonstrated achieving energy resolution per bandwidth below $\hbar$ by using nondestructive Faraday rotation probing\,\cite{palacios2022single}.
Magnetic Needle Magnetometers that feature a hard ferromagnetic needle suspended above a superconductor via the Meissner effect are anticipated to provide unparalleled sensitivity, potentially surpassing current cutting-edge devices by several orders of magnitude\,\cite{band2018dynamics,kimball2016magnetic}.
Einstein-Podolsky-Rosen entanglement of atoms are employed to enhance the sensitivity to pulsed magnetic fields for an atomic radio-frequency magnetometer\,\cite{wasilewski2010quantum,pezze2018quantum}.


\section{Exotic spin-dependent interaction searches}
\label{SecIV}

The rapid advancement of experimental techniques and technology has greatly improved the precision in controlling and measuring spin dynamics, leading to a significant enhancement in the sensitivity of spin sensors. 
This progress sets the stage for  exploring exotic spin-dependent interactions. Spin sensors play a crucial role as detectors, in tandem with astrophysical observations and particle physics detectors, in the study of these exotic interactions
\,\cite{demille2017probing,safronova2018search,budker2022quantum,antypas2022new,allanach2002snowmass,essig2022snowmass2021,semertzidis2022axion}.

\subsection{Searches for ultralight axion-like dark matter}
\label{ALPsearches}

ALPs can interact with fermion spins,
leading to energy shifts that induce a pseudomagnetic field effect—akin to the Zeeman effect\,\cite{graham2013new,budker2014proposal}.
In such scenarios, the gradient $\nabla a(\vec{r}, t)$ acts as a pseudomagnetic field [see Eq.\,\eqref{Lae} for details].
For example,
when a nuclear-spin sensor is used to measure ALP field,
the coupling constant $g_{\mathrm{a \psi}}$ is proportional to the pseudomagnetic field measured\,\cite{kimball2020overview,budker2014proposal}
\begin{equation}
    g_{\mathrm{a \psi }} \approx \frac{\delta B \hbar \gamma_i}{\sqrt{2 \hbar^3 v^2 c \rho_{\rm{d m}}}},
\end{equation}
where $g_{\mathrm{a \psi}}$ denotes the strength of
the axion coupling to the nucleon ($g_{\mathrm{aN}}$) or electron ($g_{\mathrm{ae}}$),
$\delta B$ represents the magnetic measurement sensitivity of the spin sensor (see Sec.\,\ref{Sec:sensor} for details),
and $\gamma_i$ is the nuclear gyromagnetic ratio for nuclei such as $^3$He, $^{21}$Ne, and $^{129}$Xe.
The ultimate objective of axion dark matter experiments is to detect these oscillating, nuclear-spin-dependent pseudomagnetic fields.

\begin{figure}[t]  
	\makeatletter
	\def\@captype{figure}
	\makeatother
        \centering
        \includegraphics[width=0.495\textwidth]{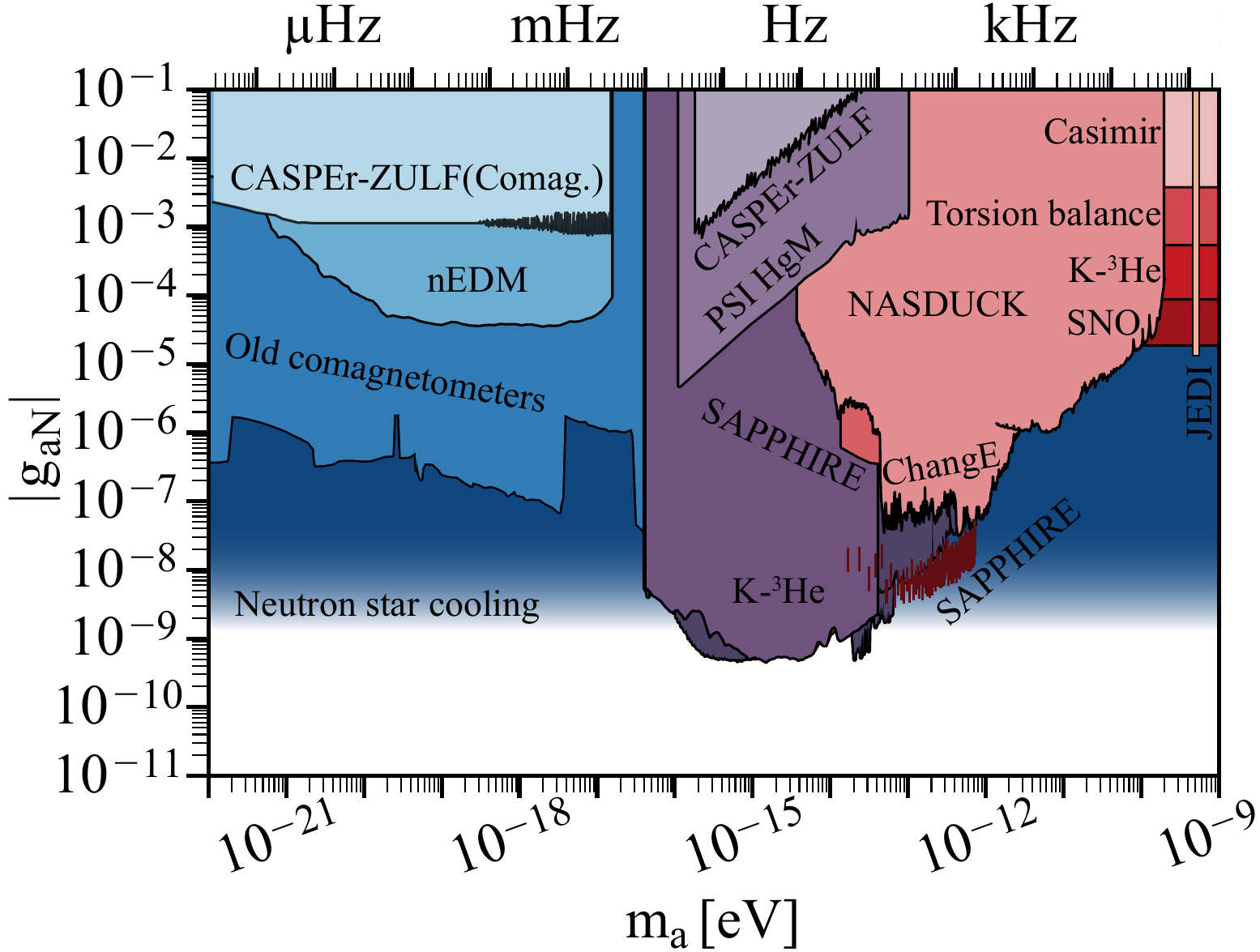}
	\caption{Constraints on axion-nucleon couplings. The laboratory constraints are from direct searches for axion-nucleon coupling and indirect searches for spin-dependent forces. Experiments at different Compton frequencies corresponds to different ALP masses $m_a$. There are important consequences that the laboratory constraints from Ref.\,\cite{jiang2021search,wei2023dark,lee2023laboratory} exceeds the astrophysical limits. The astrophysical limits are from the analysis of Ref.\,\cite{buschmann2022upper}. The figure is adapted from\,\cite{AxionLimits}.}
 \label{Figure8}
\end{figure}

A variety of nuclear-spin sensors has been utilized in the quest to detect axion-nucleon interactions (see Fig.\,\ref{Figure8}).
The neutron Electric Dipole Moment (nEDM) project utilized data from a previous $^{199}$Hg experiment\,\cite{baker2014apparatus} and a cold neutron experiment\,\cite{altarev2009test} to impose constraints on axion-nucleon couplings\,\cite{abel2017search}.
The CASPEr project proposed the use of nuclear magnetic resonance to measure axion-nucleon couplings\,\cite{budker2014proposal}.
Within the scope of the CASPEr initiatives,
the CASPEr-ZULF comagnetometer\,\cite{wu2019search} and CASPEr-ZULF sideband\,\cite{garcon2019constraints},
leveraging ZULF NMR\,\cite{kimball2020overview,ledbetter2011near},
established limits on the ALP dark matter for ALPs mass below $1.3 \times 10^{-17} \mathrm{eV}$ and between $1.8 \times 10^{-16}$ and $7.8 \times 10^{-14}\,\mathrm{eV}$.
A theoretical work analysed the old comagnetometer data from previously published data in Refs.\,\cite{vasilakis2011precision,kornack2005test,brown2011new} and placed constraints on ALP dark matter\,\cite{bloch2020axion}.

The SAPPHIRE project\,\cite{jiang2021search,jiang2022floquet,su2021search,wang2022limits,wang2023search} has pioneered and validated the utilization of long-lived nuclear spins as a pre-amplifier, significantly enhancing the detection of coherently oscillating axion-like dark matter fields by a factor of more than 100.
This amplified ALP signal is then detectable using a standard atomic magnetometer.
Employing a $^{129}$Xe spin amplifier,
SAPPHIRE has constrained the parameter space describing the coupling of
axion-like particles to nucleons in the mass range of 8.3-744.0\,feV,
at 67.5\,feV reaching $2.9\times 10^{-9}$\,GeV$^{-1}$\,\cite{jiang2021search}.
The laboratory constraints derived from five hours of data were comparable to astrophysical constraints from the cooling of supernova 1987A in multiple spectral windows\,\cite{vysotsskii1978some,raffelt2008astrophysical}.
While primarily showcased for axion-like dark matter detection, the SAPPHIRE measurements have additionally imposed constraints on the quadratic interactions between ALPs and nucleons, as well as on the interactions between dark photons and nucleons [see Eqs.\,\eqref{axialvec}, \eqref{MDMLH}, \eqref{EDMLH} in Sec.\,\ref{interaction}], surpassing the limitations set by astrophysical studies. Recently, the SAPPHIRE project has notably enhanced the amplification factor of the $^{129}$Xe spin amplifier to over 5000 and has developed a $^3$He spin amplifier with a potential amplification factor exceeding 10$^6$\,\cite{jiangdark2024}.


The Noble and Alkali Spin Detectors for Ultralight Coherent darK matter (NASDUCK) employed the $^{129}$Xe-$^{87}$Rb Floquet quantum sensor\,\cite{bloch2022new} and the $^{3}\mathrm{He}$-K SERF comagnetometer to establish constraints within the mass ranges of $4 \times 10^{-15}$-$4 \times 10^{-12}\,\mathrm{eV}$ and $1.4 \times 10^{-12}$-$2 \times 10^{-10}\,\mathrm{eV} $\,\cite{bloch2023constraints}.
The experiment searching for frequency modulation of the free spin-precession signal of Hg in 1\,$\mathrm{\mu}$T magnetic field placed new constraints of axion-nucleon couplings in the mass range of 10$^{-16}$-10$^{-13}$\,eV\,\cite{abel2023search}.
{The Jülich Electric Dipole moment Investigations (JEDI) project using in-flight spins of the beam particles in a storage ring. 
The in-plane polarization of a stored deuteron beam for a few hundred seconds. 
At resonance between the spin-precession frequency of deuterons and the ALP-induced electric dipole moment (EDM) oscillation frequency, there is an accumulation of the polarization component out of the ring plane.
The JEDI experiment imposes limits on axion-dark matter, focusing on a mass range of $4.95$-$5.02 \times 10^{-8}\,\mathrm{eV}$\,\cite{karanth2023first}.}
Moreover, recent advancements by self-compensating comagnetometers\,\cite{lee2023laboratory} and ChangE project (Coupled Hot Atom eNsembles to search for liGht dark mattEr and new physics)\,\cite{wei2023dark} have established new constraints on axion-nucleon coupling, surpassing astrophysical limits for masses between $0.4-4 \times 10^{-15}$\,eV.

Several experiments have been conducted to explore the potential couplings between axions and electrons. A ferromagnetic axion haloscope utilized a photon-magnon hybrid system linked to a quantum-limited Josephson parametric amplifier, aiming to detect axion-induced magnons\,\cite{crescini2018operation,crescini2020axion}.
The ``old comagnetometer'' constraints were derived from analyses of older comagnetometer data, under the hypothesis that axion dark matter interacts solely with electron spin\,\cite{bloch2020axion}.
However, in contexts where ALPs are hypothesized to interact with electrons, theoretical research has suggested that the interplay between the ALP field and magnetic shielding may substantially reduce the ALP-induced signal amplitudes.\,\cite{kimball2016magnetic}.
{This study conclude that magnetic shields generally do not significantly decrease sensitivity to these interactions when using nuclear spins.
However, it notes that exotic fields coupling to electron spin can produce magnetic fields within shields made of soft ferromagnetic or ferrimagnetic material, affecting the interpretation of results.
If the exotic magnetic field only couples to electron spins, interactions between the exotic field and magnetic shield will reduce the exotic signal amplitudes in each magnetometer by roughly the magnetic shielding factors of $10^6$–$10^7$.}
In a quantum nondemolition detection experiment, a ferrimagnetic sphere served as an electronic spin target alongside a superconducting qubit, both interfaced with a microwave cavity\,\cite{ikeda2022axion}.
Additionally, a fermionic interferometer, featuring two arms with orthogonal spin resonances, was employed to investigate the axion-induced precession of an electron spin resonance\,\cite{crescini2023fermionic}.


\subsection{Searches for exotic spin-dependent forces}

The theoretical study by Dobrescu et al. proposed fifteen potential exotic spin-dependent forces\,\cite{Dobrescu:2006au},
which can be categorized into two primary categories:
spin-mass forces, represented by ${V}_{4+5},{V}_{9+10},{V}_{12+13}$ and spin-spin forces, represented by ${V}_{2},{V}_{3},{V}_{6+7},{V}_{8},{V}_{11},{V}_{14},{V}_{15},{V}_{16}$.
Current laboratory searches for these forces typically involve two distinct particle ensembles: a polarized fermion ensemble serving as a ``spin sensor" and another ensemble acting as a ``mass source" with unpolarized fermions\,\cite{zhang2023search,Wu2023,wu2022experimental,su2021search,wei2022constraints,feng2022search,aggarwal2022characterization,jiao2021experimental,lee2018improved,kim2019experimental,rong2018constraints,xiao2023femtotesla,xiao2024exotic,wineland1991search,youdin1996limits,kimball2017constraints,chu2013laboratory,bulatowicz2013laboratory,wu2023new,clayburn2023using}, or as a ``spin source" utilizing polarized fermions\,\cite{almasi2020new,vasilakis2009limits,glenday2008limits,ji2023constraints,rong2018searching,wang2022limits,wang2023search,ji2018new,kotler2015constraints,hunter2014using,hunter2013using}.
These experiments aim to search for the exotic interactions between the spin or mass source and the spin sensor, potentially inducing a pseudomagnetic field on the spin sensor.
A wide variety of spin sensors have been utilized to detect these induced pseudomagnetic fields.
Techniques and devices such as SERF magnetometers, spin amplifiers, NMR techniques, comagnetometers, spin masers, NV diamond sensors, particle beam experiments, ion traps, and torsion balances have all been employed in the pursuit of detecting exotic spin-dependent forces\,\cite{lee2018improved,almasi2020new,vasilakis2009limits,glenday2008limits,ji2023constraints,rong2018constraints,rong2018searching,jiao2021experimental,su2021search,wang2022limits,wang2023search,feng2022search,zhang2023search,ji2018new,wei2022constraints}.

For example, the spin-mass force ${V}_{9+10}$ and spin-spin force ${V}_{3}$ can be mediated by axions and generate pseudomagnetic fields $\vec{B}_a$ on the spin sensor $\vec{B}_a \cdot \vec{{\sigma}}_{1}=-{{V}}/{\mu_{i}}$.
{This phenomenon can be mathematically described as follows
\begin{equation}
   \delta B= \left(\frac{g_p g_p}{4}, g_sg_p \right){N |\vec{B}_a|},
\end{equation}
where $g_p$ ($g_s$) represents the pseudoscalar (scalar) coupling constants, $N$ denotes the number of polarized (for ${V}_3$) or unpolarized (for ${V}_{9+10}$) fermions in the source,
$\vec{B}_a$ is determined by the geometric structure between the source and sensor with the assumption ($g_i g_i=1$), and $\mu_{i}$ represents the magnetic moment of the sensor spin.}
To enhance the projected sensitivity to the coupling strength, several strategies can be employed, such as increasing the number ($N$) of fermions in the source, enhancing the sensitivity ($\delta B$) to magnetic fields and reducing the distance between the source and sensor. The principal obstacle in conducting such measurements is the identification and mitigation of systematic errors, particularly in distinguishing the effects of the pseudomagnetic field from those of classical electromagnetic interactions.

\begin{figure}[t]  
	\makeatletter
	\def\@captype{figure}
	\makeatother
        \centering
        \includegraphics[width=0.49\textwidth]{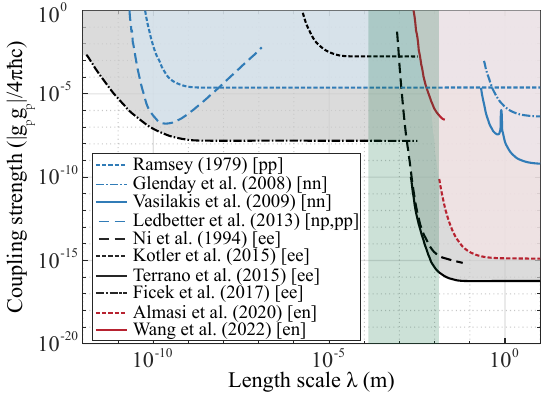}
	\caption{Laboratory constraints on pseudoscalar $g_p g_p$ couplings as the function of axion Compton wavelength $\lambda$. Experiments at different length scales measure interaction ranges corresponding to different axion Compton wavelengths $\lambda$ and thus different axion masses $m_a$. The shaded green region represents the axion window. }
 \label{Figure9}
\end{figure}

A wide range of experiments, utilizing various types of spin sensors, have been undertaken to probe exotic spin-dependent forces.
The effective range of forces that can be investigated is generally constrained by the dimensions of the sensor used.
For the force range shorter than $10^{-4}$\,m, the constraints are mostly set by torsion balance\,\cite{ni1994search}, atomic or molecular spectroscopy\,\cite{ledbetter2013constraints}, molecular beam\,\cite{ramsey1979tensor},
hyperpolarized $^{3}$He\,\cite{petukhov2010polarized,guigue2015constraining} and NV diamond\,\cite{rong2018searching,rong2018constraints,jiao2021experimental,liang2022,Wu2023}.
For the force range above $10^{-4}$\,m, the constraints are mostly set by atomic comagnetometer\,\cite{wei2022constraints,almasi2020new,lee2018improved,hunter2013using,hunter2014using,venema1992search,vasilakis2009limits,glenday2008limits,youdin1996limits,kimball2017constraints,chu2013laboratory,zhang2023search,feng2022search,bulatowicz2013laboratory}, spin amplifier\,\cite{su2021search,wang2022limits,wang2023search}, NMR techniques\,\cite{chu2013laboratory}, ion trap\,\cite{wineland1991search,kotler2015constraints}, torsion balance\,\cite{terrano2015short,hoedl2011improved,heckel2008preferred}, methods measuring the variation of magnetization in solids\,\cite{crescini2017improved,ni1999search}.
This diverse toolkit of experimental techniques underscores the multifaceted approach required to explore the potential existence and characteristics of exotic spin-dependent forces, each with its unique advantages and limitations based on the force range and experimental setup.

Experimental constraints on $g_s g_p$ predominantly utilized comagnetometer\,\cite{tullney2013constraints,bulatowicz2013laboratory}, NMR techniques\,\cite{chu2013laboratory}, torsion balance\,\,\cite{hoedl2011improved,terrano2015short}.
In comagnetometer experiments, the spin precession frequencies of co-located $^3 \mathrm{He}$ and $^{129} \mathrm{Xe}$ gases were measured using a multichannel SQUID\,\cite{tullney2013constraints},
employing a BGO crystal $\left(\mathrm{Bi}_4 \mathrm{Ge}_3 \mathrm{O}_{12}\right)$ as the mass source due to its high nucleon number density and low conductivity.
This BGO crystal was adjustably positioned between approximately 2 and 200\,mm from the sensor cell via a compressed-air piston.
Another comagnetometer approach utilized a ${ }^{129} \mathrm{Xe}$-${ }^{131} \mathrm{Xe}$ comagnetometer to investigate monopole-dipole interactions,
deploying a zirconia rod near the NMR cell as a movable mass\,\cite{bulatowicz2013laboratory}.
In NMR experiments, the spin-precession frequency shift of polarized $^3$He was measured
when a mass source—either a ceramic block or a liquid mixture containing roughly $\approx 1 \%$ $\mathrm{MnCl}_2$ and pure water—was transitioned from $5 \mathrm{~cm}$ to $10 \mu \mathrm{m}$ from the spin sensor using a stepping motor.
These specific masses were selected for their nucleon densities, minimal magnetic impurities and susceptibilities, and negligible impact on NMR measurements.

Experimental constraints on $g_p g_p$ have been extensively established through a variety of methods (see Fig.\,\ref{Figure9}),
including molecule beam experiment for $g_p^p g_p^p$\,\cite{ramsey1979tensor},
comagnetometer for $g_p^e g_p^n$, $g_p^n g_p^n$\,\cite{almasi2020new,vasilakis2009limits}, spectroscopic measurement for $g_p^e g_p^e$\,\cite{ficek2017constraints},
ion trap for $g_p^e g_p^e$\,\cite{kotler2015constraints}, and spin amplifier for $g_p^e g_p^n$\,\cite{wang2022limits}.
For example,
ion trap experiments involved comparing the magnetic dipole-dipole interaction between two trapped $^{88}\mathrm{Sr}^{+}$ions\,\cite{kimball2017constraints} against theoretical values\,\cite{kotler2015constraints}, thereby establishing new constraints on $g_p^e g_p^e$.
In comagnetometer studies,
a $^{3}$He-K comagnetometer was deployed to detect pseudomagnetic fields produced by a hyperpolarized $^{3}$He source\,\cite{vasilakis2009limits},
with the nuclear-spin direction of the $^3$He-gas spin source being inverted at a 0.18\,Hz frequency through adiabatic fast passage.
In the context of spin amplifier experiments, the search for exotic forces between neutron and electron was conducted using a spin amplifier coupled with an optically pumped $^{87}$Rb spin source\,\cite{wang2022limits},
where the polarization of $^{87}$Rb atomic spins was periodically modulated by blocking the pump beam at a frequency $\approx 10.00$\,Hz with a chopper.

Recently, a number of theoretical frameworks—including high-temperature lattice QCD\,\cite{borsanyi2016calculation},
the SM Axion Seesaw Higgs portal inflation (SMASH) model\,\cite{ballesteros2017unifying},
and axion-string networks\,\cite{klaer2017dark,turner1990windows,youdin1996limits,arvanitaki2014resonantly,yang2019probe}—highlight a specific mass range known as the ``axion window" (10\,$\mu$eV to 1\,meV).
This range is considered to be one of the most probable areas in the parameter space where axions might be found.
Investigations into spin-dependent forces offer a promising avenue for probing axions within the ``axion window" without the necessity to scan the resonance frequency. This approach is particularly significant given the challenges associated with direct experimental searches for axion dark matter within this mass range, which typically necessitate the use of strong magnetic fields—a requirement that can be technically demanding and resource-intensive.
To effectively explore this range, the spatial distance between the sensor and the source in experiments needs to be maintained at less than several centimeters, emphasizing the delicate nature of these measurements and the precision required in experimental setups to detect such elusive forces.
Within this context, for example, the SAPPHIRE project used a $^{129}$Xe amplifier as spin sensor to amplify the pseudomagnetic fields emanating from exotic interactions between $^{87}$Rb spin source and $^{129}$Xe amplifier\,\cite{wang2022limits}.
The spatial arrangement was meticulously designed, positioning the spin source 39\,mm away from the center of the spin-based amplifier.
This approach has enabled SAPPHIRE to enhance the signal from pseudomagnetic fields by more than a factor of 40. Leveraging this spin-based amplifier, SAPPHIRE has set a direct upper limit on the magnitude of $g_p^e g_p^n$ for pseudoscalars and ventured into previously uncharted parameter space for axion masses ranging from 0.03\,meV to 1\,meV within the axion window\,\cite{wang2022limits}.

Single NV centers in diamond have emerged as a solid-state spin quantum sensor to search for exotic spin-dependent interactions at the micrometer scale, exploiting the ability to enable close proximity between the sensor and the source. The single electron spin of a near-surface NV center was used as the quantum sensor, while a fused-silica half-sphere lens was employed as the source of the moving nucleons to search for exotic spin-mass interactions. Combined with an AFM setup, the force range can be focused within micrometers\,\cite{rong2018searching}. 
In another study, a single crystal of p-terphenyl doped pentacene-d14 under laser pumping provided the source of polarized electron spins. Constraints on the exotic electron-electron coupling were set based on the measurement of polarization signal using single NV sensors\,\cite{rong2018constraints}. 
Afterward, several exotic spin-mass interactions were scrutinized further using the ensemble-NV-magnetometer with improved sensitivity\,\cite{liang2022,Wu2022c}. 
Recently, new constraints on exotic spin-spin-velocity-dependent interactions between electrons have been established using two NV ensembles, one as the spin sensor and the other as the spin source.
The series of experiments fully demonstrates the potential of NV centers to explore exotic interactions exploiting the compact, flexible, and sensitive features of solid-state spins\,\cite{huang2024new}.

Exotic spin-dependent forces are characterized by dimensionless coupling constants and exhibit a rapid decay in strength with increasing distance between interacting fermions (see Sec.\,\ref{forcesearches} for details)\,\cite{Fadeev2019,Dobrescu:2006au}.
Due to the dual vertices involved in the scattering of two fermions, the sensitivity to the coupling constant is attenuated by two orders of the small coupling constant, either $g_{\mathrm{aN}}$ or $g_{\mathrm{ae}}$.
However, the searches for exotic spin-dependent forces enables the exploration of significantly heavier dark matter candidates, such as axions and ALPs, with masses reaching up to 1\,meV (equivalent to frequencies around 100\,GHz).
Achieving such detection through direct measurement methods (see Sec.\,\ref{ALPsearches}) poses a considerable challenge, as identifying axions at 100\,GHz would necessitate magnetic fields exceeding 100\,T.
An additional advantage of these indirect measurement strategies is their capability to span a broad mass range (e.g. $10^{-22}$-$10^{-3}$\,eV) within a single experimental setup, substantially streamlining the detection process relative to methods that require scanning across resonance frequencies.
Furthermore, investigations into the existence of spin-1 bosons,
such as Z$'$ and $\gamma'$ bosons\,\cite{Dobrescu:2006au,Fadeev2019},
frequently focuses on the investigation of spin-dependent forces.

\subsection{Searches for EDMs}

The quest to discover fundamental EDMs has been in progress since Purcell and Ramsey initiated neutron EDM beam experiments in 1949\,\cite{smith1957experimental}.
The motivation behind EDM research primarily stems from the observation of Parity- or Time-reversal symmetry violation (PT-violation), which suggests the existence of hypothetical particles beyond the SM, thereby linking to the baryogenesis problem\,\cite{cairncross2019atoms,safronova2018search,chupp2019electric}.
An atomic or molecular EDM may originate from the EDMs of its constituent electrons or nucleons, or through other T-violating interactions among the constituent particles\,\cite{fleig2018model,chupp1988precision}.
The presence of the permanent EDMs induces a Zeeman-like energy shift that varies linearly with an applied electric field $\vec{E}$.
Numerous experiments have adopted the Ramsey interferometry to measure the frequency shift caused by the electric fields\,\cite{roussy2023improved,acme2018improved,ramsey1990experiments,ramsey1986neutron,abel2020measurement}
\begin{equation}
   |d| \approx \frac{\hbar |\delta \omega|}{|\vec{E}|}.
\end{equation}
The current measurement precision of EDM experiment can achieve the level of $10\,\mu$Hz\,\cite{roussy2023improved}.

 
Currently, the leading experiments dedicated to probing the EDMs can be categorized into three primary systems\,\cite{chupp2019electric,kuchler2019searches,cairncross2019atoms,acme2018improved,roussy2023improved,abel2020measurement,shmakova2023electric}: fundamental particles and nucleons, diamagnetic system, and paramagnetic system (see Table\,\ref{EDMtable}).
Diamagnetic systems, characterized by possessing zero electron spin but a nonzero nuclear spin, primarily derive their EDMs from the nuclear Schiff moment. This moment predominantly encompasses contributions from the EDMs of neutrons and protons as well as from isoscalar and isovector pion-nucleon interactions. Additionally, diamagnetic EDMs may emerge from scalar and tensor electron-nucleon interactions. Notably, diamagnetic atoms and molecules are theoretically sensitive to 11 out of the 13 BSM CP-violating effective operators\,\cite{chupp2019electric,cairncross2019atoms,fleig2018model,ginges2004violations}, which are anticipated to significantly influence EDMs.
Conversely, paramagnetic systems, which feature one or more unpaired electron spins, have permanent EDMs of paramagnetic atoms and molecules mainly originating from the electron EDM, denoted as $d_e$, and the scalar electron-nucleon coupling.

\begin{table}[t]
\newcommand{\tabincell}[2]{\begin{tabular}{@{}#1@{}}#2\end{tabular}}
\begin{ruledtabular}
\caption {Summary of published experimental upper limits and Standard Model (SM) predictions for several systems of fundamental particles and atoms used in electric dipole moment (EDM) searches. Experimental results are far beyond SM predictions\,\cite{ellis1989theory,chupp2015electric,kuchler2019searches} but are within reach of beyond SM theories to constrain model parameters. The $^{199}$Hg and $^{129}$Xe EDMs searches were conducted based on comagnetometry\,\cite{graner2016reduced,abel2020measurement}. The $^{225}$Ra EDM, nEDM and eEDM searches were conducted based on Ramsey interferometry\,\cite{abel2020measurement,roussy2023improved,regan2002new,bishof2016improved,hudson2011improved,acme2018improved}.} 
\label{EDMtable}
\renewcommand{\arraystretch}{1.2}
\begin{tabular}{c c c c}   
Spin system & Exp. ($e$ cm) & SM Pred. ($e$ cm) & Ref. \\
\midrule
Ultracold neutron & $1.8 \times 10^{-26}$ {(90\,\%)} &$\sim 10^{-32}$-$10^{-31}$&\,\cite{abel2020measurement}\\
\midrule
${ }^{199}\mathrm{Hg}$&$7.4 \times 10^{-30}$ {(95\,\%)}&$\sim 10^{-34}$ &\,\cite{graner2016reduced}\\[0.15cm]
${ }^{129}$Xe &$1.4 \times 10^{-27} $ {(95\,\%)} & $\sim 10^{-34}$&\,\cite{sachdeva2019new}\\[0.15cm]
${ }^{225}$Ra & $1.4 \times 10^{-24}$ {(95\,\%)}&$-$&\cite{bishof2016improved} \\[0.15cm]
$^{171}$Yb &$1.5 \times 10^{-26}$ {(95\,\%)} &$-$&\cite{zheng2022measurement}\\
\midrule
Electron ($\mathrm{HfF}^{+}$) & $
4.1 \times 10^{-30}$ {(90\,\%)}& $\sim 10^{-38}$ &\,\cite{roussy2023improved}\\[0.15cm]
Electron (ThO)& $1.1 \times 10^{-29}$ {(90\,\%)}&$\sim 10^{-38}$&\,\cite{acme2018improved}\\[0.15cm]
Electron (YbF)& $1.1 \times 10^{-27}$ {(90\,\%)}&$\sim 10^{-38}$&\,\cite{hudson2011improved}\\[0.15cm]
Electron (Tl) &$1.6 \times 10^{-27}$ {(90\,\%)} &$\sim 10^{-38}$&\,\cite{regan2002new}\\[0.05cm]
\end{tabular} 
\end{ruledtabular}
\end{table}

{The searches for EDMs of fundamental particles encompass experiments targeting neutrons, protons, muons, protons, deuterons, and helions\,\cite{abel2020measurement,ayres2021design,serebrov2017ucn,ito2018performance,shmakova2023electric,rathmann2013search}}.
Due to the significant systematic errors caused by beam divergence and motional $(\vec{v} \times \vec{E})$ effects, nEDM experiments post-1980s have exhibited a preference for utilizing ultracold neutron beams\,\cite{smith1990search,baker2006improved,abel2020measurement}. 
This shift to ultracold neutron experiments has led to remarkable advances in sensitivity, facilitated by considerably longer interaction times compared to early neutron beam experiments.
The most recent findings from the nEDM experiment were reported by the Paul Scherrer Institute\,\cite{abel2020measurement}.
A notable aspect of this experiment was the deployment of a $^{199}$Hg comagnetometer and an array of optically pumped cesium vapor magnetometers to cancel and correct for magnetic-field changes.
The measured value of the nEDM from this experiment is $\left|d_n\right|<$ $1.8 \times 10^{-26}$$\,e$\,cm $(90\,\%$ C.L.),
with an electric field of $11\,\mathrm{kV} / \mathrm{cm}$ being applied. 
{Additionally,
the JEDI storage-rings experiment is aimed to detect the effect of the EDM on the spin motion of a charged particle in the presence of magnetic and electric fields\,\cite{karanth2023first}.}
The first direct measurement of the deuteron EDM at COSY finished and the analysis is currently ongoing\,\cite{shmakova2023electric}.

Various diamagnetic systems are employed for EDMs searches, including $^{199} \mathrm{Hg}$, $^{225}\mathrm{Ra}$, $^{129}\mathrm{Xe}$, $\mathrm{Rn}$ and TIF\,\cite{romalis2001new,griffith2009improved,graner2016reduced,bishof2016improved,sachdeva2019new,pen2014design,cho1991search}.
Utilizing diamagnetic atoms offers several advantages, such as high polarization, extended spin-coherence lifetimes, and compatibility with room temperature operations.
A sequence of experiments with $^{199}$Hg atoms have been carried out, employing comagnetometry to measure the spin precession frequency\,\cite{romalis2001new,griffith2009improved,graner2016reduced}.
In these experiments,
$^{199}$Hg contained within four stacked, coated cells was directly excited and observed using a 254\,nm laser, positioned perpendicularly to the magnetic and electric fields. The arrangement ensured that the outer two cells, which were devoid of an electric field, and the inner two cells, which had electric fields in opposing directions, could facilitate the identification of an EDM signal through the difference in free-precession frequencies between the two inner cells.
The established upper limit for the $^{199}$Hg EDM is $\left|d_{\mathrm{Hg}}\right|<7.4 \times 10^{-30} $\,$e$\,cm (95\% C.L.).
The $^{129}$Xe EDM experiment typically incorporates a $^{129}$Xe-$^{3}$He comagnetometer\,\cite{sachdeva2019new,allmendinger2019measurement},
leveraging their distinctly different sensitivities to the Schiff moment and other P-odd and T-odd interactions, yet similar susceptibility to magnetic-field influences.
The most recent limit of $^{129}$Xe EDM measurement utilized SQUID Detection to monitor the free precession of the two spins\,\cite{sachdeva2019new},
establishing an upper limit for the Xe EDM at $\left|d_{\mathrm{Xe}}\right|<$ $1.4 \times 10^{-27}$\,$e$\,cm (95\% C.L.).
The sensitivity of $^{225}\mathrm{Ra}$ EDM to the P-odd and T-odd pion-nucleon couplings could surpass that of $^{199}$Hg by 2–3 orders of magnitude.
Experiments targeting $^{225}\mathrm{Ra}$ are actively being conducted, with the current upper limit set at $1.4\times10^{-24}$\,$e$\,cm (95\% C.L.)\,\cite{bishof2016improved}.
Because the radioactivity and rarity of $^{225}\mathrm{Ra}$ cause considerable
difficulties in the development of the cold-atom optical dipole trap method for the EDM measurement, current experiments set limit on the atomic EDM of $^{171}$Yb $\left|d_\mathrm{Yb}\right|<1.5 \times 10^{-26} e \mathrm{~cm}$ (95\% C.L.)\,\cite{zheng2022measurement}.

In the exploration of paramagnetic systems for EDMs, early experiments used atomic beams of paramagnetic atoms such as Cs and Tl.
However, the sensitivity of these experiments was constrained by issues such as wide linewidths, count-rate limitations, and significant coupling of the large magnetic moments to motional magnetic fields.
Since the 2000s, the search for the electron EDM (eEDM) has shifted towards employing paramagnetic polar molecules composed of one light and one heavy atom, for example, $\mathrm{YbF}$, ThO, and $\mathrm{Hf} \mathrm{F}^{+}$\,\cite{roussy2023improved,acme2018improved,hudson2011improved}.
These molecules are subject to a strong interatomic electric field, typically on the order of $\mathrm{GV} /\mathrm{cm}$, which significantly exceeds what can be achieved with laboratory electric fields.
The most recent advancement in measuring the eEDM involved the use of $\mathrm{HfF}^{+}$ molecular ions\,\cite{roussy2023improved}.
In this experimental configuration,
an electric field of approximately $58\,\mathrm{V}/\mathrm{cm}$ was applied,
resulting in an effective electric field of approximately $ 23\,\rm{GV}/\rm{cm}$.
The application of a minor magnetic field influenced the orientation of the valence electron's spin, causing it to align or anti-align with the effective electric field. 
A coherent superposition of these two spin states was established, and the energy difference between them was determined through Ramsey spectroscopy.
This methodology led to an eEDM value of $|d_e|<4.1 \times 10^{-30}$\,$e$\,cm $(90 \%$ C.L.)\,\cite{roussy2023improved}.

\subsection{Searches for spin-gravity interactions}

{Direct comparison experiments of the gravitational acceleration between different spin states or between a spin polarized object and a sample of randomly polarized spins in the presence of a magnetic field have been discussed since the 1950s\,\cite{mcreynolds1951gravitational,dabbs1965gravitational,witteborn1967experimental}.}
In this type of experiments,
the current measurement for Eötvös parameters based on free fall of cold atoms achieved precision levels of $10^{-7}$\,\cite{duan2016test,tarallo2014test}.
The second type of experiments related to spin-dependent Eötvös parameters involves comparing the energy associated with anomalous spin couplings to gravitational potential energy to measure the effects of spin in the solar gravitational field, with measurement precision levels of $10^{-20}$\,\cite{venema1992search,zhang2023search} (see Table\,\ref{tab:my_label3}).

\begin{table}[b]
\newcommand{\tabincell}[2]{\begin{tabular}{@{}#1@{}}#2\end{tabular}}
\begin{ruledtabular}
\caption {Constraints (95\% C.L.) on the energy difference $\hbar\left|A_{\mathrm{sg}}\right|$ due to the spin-gravity coupling and the Eötvös parameter $\eta_{\mathrm{sg}}$.} 
\label{tab:my_label3}
\renewcommand{\arraystretch}{1.2}
\begin{tabular}{c c c c c}  
System& \multirow{2}{*}{\centering Spin} & \multirow{2}{*}{\centering $\hbar\left|A_{\mathrm{sg}}\right|(\mathrm{eV})$} & \multirow{2}{*}{\centering $\left|\eta_{\mathrm{sg}}\right|$}&\multirow{2}{*}{\centering Ref.}\\
(Sensor)&  & & &\\
\midrule
AlNiCo-SmCo${ }_5$ &\multirow{2}{*}{\centering Electron} &\multirow{2}{*}{\centering $2.2 \times 10^{-19}$}& \multirow{2}{*}{\centering $1.2 \times 10^{-15}$}&\multirow{2}{*}{\centering \cite{heckel2008preferred}}\\
(Torsion balance) & & & &\\[0.15cm]
${ }^{85}\mathrm{Rb}$-$^{87} \mathrm{Rb}$ &  \multirow{2}{*}{\centering Proton}&  \multirow{2}{*}{\centering $3.4 \times 10^{-18}$}&  \multirow{2}{*}{\centering $1.1 \times 10^{-17}$}& \multirow{2}{*}{\centering \cite{kimball2017constraints}}\\
\,\,\,\,(Ion trap) & & & &\\[0.15cm]
${ }^9 \mathrm{Be}^{+}$ & \multirow{2}{*}{\centering Neutron} & \multirow{2}{*}{\centering $1.7 \times 10^{-19}$} & \multirow{2}{*}{\centering $5.4 \times 10^{-19}$}& \multirow{2}{*}{\centering \cite{wineland1991search}} \\
(Ion trap) & & & &\\[0.15cm]
${ }^{199} \mathrm{Hg}$-$^{201}\mathrm{Hg}$ & \multirow{2}{*}{\centering Neutron} & \multirow{2}{*}{\centering $9.1 \times 10^{-21}$} & \multirow{2}{*}{\centering $2.9 \times 10^{-20}$}&\multirow{2}{*}{\centering \cite{venema1992search}} \\
(Comagnetometer) & & & &\\[0.15cm]
${ }^{129} \mathrm{Xe}$-$^{131} \mathrm{Xe}$ & \multirow{2}{*}{\centering Neutron} & \multirow{2}{*}{\centering $5.3 \times 10^{-22}$} & \multirow{2}{*}{\centering $1.7 \times 10^{-21}$} & \multirow{2}{*}{\centering \cite{zhang2023search}}\\
(Comagnetometer) & & & &\\
\end{tabular} 
\end{ruledtabular}
\end{table}


Investigations into atomic energy shifts associated with the reorientation of the quantization axis in relation to Earth's gravitational field have utilized various experimental setups, including torsion pendulums,
$^9 \mathrm{Be}^{+}$ions confined in Penning traps\,\cite{wineland1991search},
and comagnetometers incorporating diverse isotopic pairs,
including $^{85} \mathrm{Rb}-^{87} \mathrm{Rb}$\,\cite{kimball2017constraints},
$^{199} \mathrm{Hg}-^{201} \mathrm{Hg}$\,\cite{venema1992search},
and $^{129}$Xe-$^{131}$Xe\,\cite{zhang2023search}.
In these experiments, the fundamental approach involves measuring the differential spin-precession frequencies of various species induced by the gravitational field. 
The energy difference can be derived from the measured frequency difference $\hbar|A_{\rm{sg}}|=\hbar |\delta \omega|$ caused by the spin gravity interactions.
Based on Eq.\,\eqref{SG}, the corresponding Eötvös parameter $\eta_{\mathrm{sg}}$ is 
\begin{equation}
   \left|\eta_{\mathrm{sg}}\right|\approx \frac{2 \hbar\left|\delta \omega \right|}{m_{n,e} g\left(r_E\right) r_E},
\end{equation}
{where $r_E$ is the radius of earth for the ground-state experiment and $g(r_E)$ is the gravitational acceleration at $r_E$.}
For instance, constraints on spin-gravity interactions were explored using ${ }^{129} \mathrm{Xe}$ and ${ }^{131} \mathrm{Xe}$,
which were positioned on a dual rotation table and a tilt table setup\,\cite{zhang2023search}.
By alternating the orientation of the bias field to be parallel and antiparallel to the direction of Earth's rotation, frequency measurements were achieved with a precision level of 65\,nHz. This result further established an upper boundary of $2.7\,\mathrm{fm}$ for the offset between the neutron's center of mass and its center of gravity.


\section{Network of distributed spin sensors}
\label{SecV}

\begin{figure}[t]  
	\makeatletter
	\def\@captype{figure}
	\makeatother
        \centering
	\includegraphics[width=0.45\textwidth]{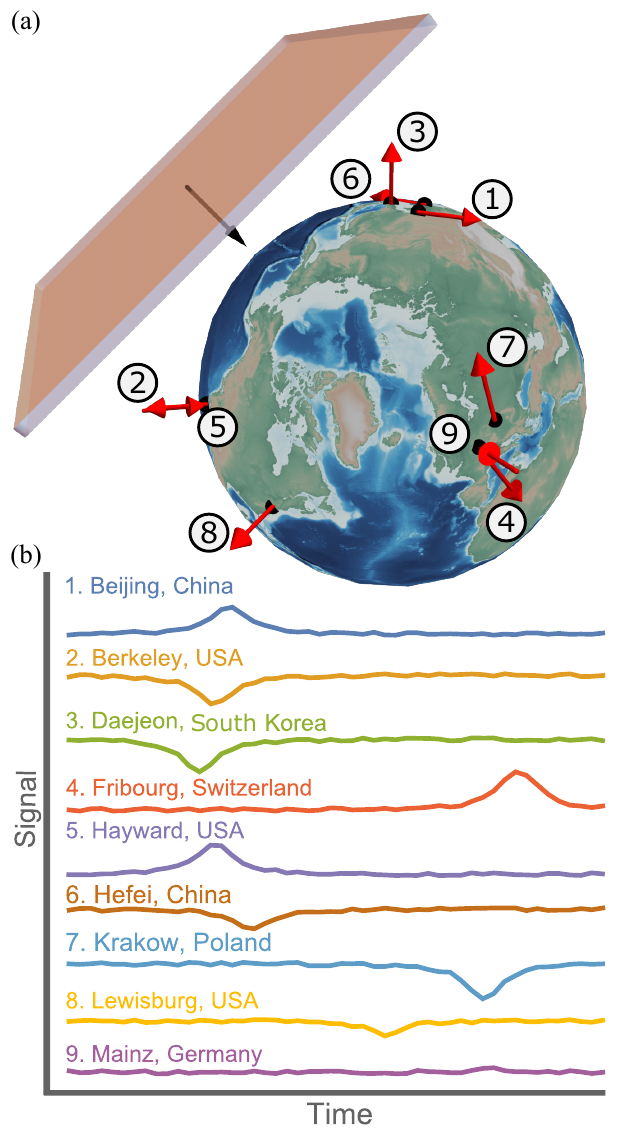}
	\caption{Global Network of Optical Magnetometers for Exotic physics (GNOME) composed of optically pumped magnetometers. (a) An image depicts the Earth alongside the location and orientation of the sensitive axes of the GNOME magnetometers. The positions and sensitive axes are indicated by red arrows. (b) A figure illustrates a simulation of the signals anticipated from a domain-wall crossing event, as detected by the various magnetometers within the network. The figure is from Ref.\,\cite{Afach:2021pfd}.}
 \label{GNOME}
\end{figure}

\begin{figure*}[t]  
	\makeatletter
	\def\@captype{figure}
	\makeatother
        \centering
	\includegraphics[width=0.9\textwidth]{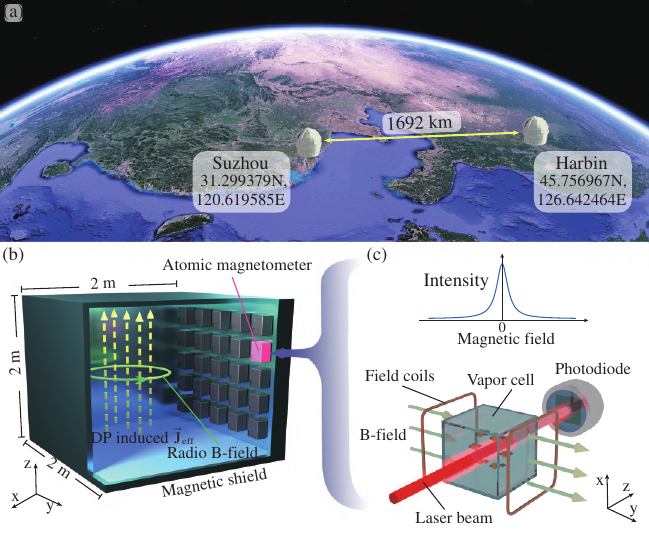}
	\caption{Quantum sensor network. (a) The atomic magnetometer network (AMAILS) composed of 15 spin-exchange relaxation-free (SERF) magnetometers, which are situated in two separate shield rooms in Suzhou and Harbin, China. Each room is made of five-layer mu-metal and its innermost layer has the dimension of $2\times 2\times 2$\,cm$^3$. (b) Schematic for dark photon dark matter induced magnetic field inside a shield room. All SERF magnetometers are installed on the surface of the shield room, where the signal induced by dark matter is at its peak. (c) Schematic of atomic magnetometer based on zero-field resonance. The figure is from Ref.\,\cite{jiang2023search}.}
 \label{AMAILS}
\end{figure*}

Networks of distributed spin sensors have recently attracted an extensive attention in exotic spin-dependent interactions searches, in particular they could reduce noise in single detector to thus achieve high detection sensitivity.
Moreover, sensor networks show unique advantages in some certain models beyond the SM.
For example,
instead of a uniform flux, ultralight bosonic fields can form stable macroscopic field configurations known as topological dark matter structures such as cosmic strings, domain walls, and textures, arising from symmetry breakdowns or self-interactions\,\cite{arvanitaki2011exploring,graham2018spin,baryakhtar2021black,kimball2018searching}.
Moreover, intense bursts of exotic ultralight bosonic fields can be produced during high-energy astrophysical phenomena\,\cite{dailey2020quantum}.
In these scenarios, terrestrial detectors observe transient events when these ultralight bosonic fields traverse the Earth.
Several challenges are involved in detecting these transient fields:
Firstly, the desired signal may be obscured in long-term averages.
Secondly, distinguishing the sought-after transient signal from transient spurious noise can be challenging.
Thirdly, measuring the transient signal from a source becomes difficult when the sensor's sensitivity axis is orthogonal to the field direction.
A network of geographically distributed sensors is well-suited for detecting these transient fields\,\cite{pospelov2013detecting,Derevianko:2013oaa,Derevianko:2016vpm,masia2020analysis,Chen:2021bdr}.
By cross-correlating data from different sensors, local noise effects can be mitigated, enabling the differentiation between exotic physics and conventional standard-model phenomena.
This approach leads to enhanced rejection of false positives, improved sensitivity, and increased confidence in identifying the dark matter origin of the signal being sought.
These networks have been demonstrated by spin sensors such as GNOME\,\cite{Afach:2021pfd,afach2024can},
and the atomic magnetometer network (AMAILS)\,\cite{jiang2023search},
as well as non-spin sensors like the atomic and optical clock network\,\cite{roberts2017search,wcislo2018new} and the SuperMAG network\,\cite{Fedderke:2021rrm}.


The GNOME network [Fig.\,\ref{GNOME}(a)], a global initiative comprising over a dozen optical atomic magnetometers stationed across Europe, North America, Asia, the Middle East, and Australia, operates in synchronization with GPS to detect pseudomagnetic fields generated by ALP domain wall dark matter\,\cite{Afach:2021pfd,afach2024can,Afach:2018eze}.
Each optical atomic magnetometer in the GNOME network is based on optically pumped alkali-metal spins and shows a magnetic sensitivity of approximately $100\,\mathrm{fT}$/Hz$^{1/2}$ over a bandwidth of approximately $100\,\mathrm{Hz}$.
Figure\,\ref{GNOME}(b) shows a simulation of the signals
expected to be observed from a ALP domain wall crossing at the different GNOME magnetometers\,\cite{Afach:2021pfd}.
The extensive geographic distribution of magnetometers allows GNOME to attain high spatial resolution, functioning as an ``exotic physics telescope" with a baseline akin to Earth's diameter.
The GNOME is sensitive to ALPs with masses between 10$^{-17}$-10$^{-9}$\,eV.
Using GNOME to investigate transient couplings between atomic spins and ALP domain wall dark matter, the initial GNOME iteration explores the ALP parameter space up to $f_{\text {int }} \approx 4 \times 10^5\,\mathrm{GeV}$\,\cite{Afach:2021pfd}, which can be furthered improved by Advanced GNOME using noble-gas
comagnetometers\,\cite{afach2024can}.



There are other well-motivated dark matter candidates, such as dark photons\,\cite{caputo2021dark}.
They behave as effective currents when coupled with Standard-Model photons  via kinetic mixing.
When situated in electromagnetic shielded rooms, oscillating magnetic fields are generated with the maximum field strength proportional to the shield’s size.
A recent search for dark photons using synchronized Atomic Magnetometer Arrays In Large Shields (AMAILS) has been reported\,\cite{jiang2023search},
involving a deployment of 15 SERF atomic magnetometers.
As shown in Fig.\,\ref{AMAILS}(a),
such magnetometers are situated in two separate electromagnetic shielded rooms in Harbin and Suzhou, China, with a distance of about 1700\,km between the locations, and are synchronized with GPS.
Both shielded rooms are made of five-layer mu-metal and their innermost layer is cuboid in shape with the dimension of $2\times 2\times 2$~m$^3$.
The magnetometer captures dark photon-induced current,
which subsequently produces a magnetic field running tangentially parallel to the walls in a horizontal direction.
To detect the maximum magnetic field generated by dark photons, all magnetometers are mounted on the walls of shielded rooms [Fig.\,\ref{AMAILS}(b)] with a magnetic field sensitivity of approximately $\approx 15\,\mathrm{fT}/\mathrm{Hz}^{1/2}$.
In the absence of a magnetic field, the spin magnetic moments align with the pump beam, leading to maximized laser light transmission to the photodiode. However, when a magnetic field perpendicular to the beam is introduced, it induces Larmor precession, rotating the magnetic moments out of alignment. This rotation results in a noticeable reduction in light transmission [Fig.\,\ref{AMAILS}(c)].
This phenomenon generates a zero-field resonance, serving as an exceedingly sensitive indicator of magnetic fields.
By correlating the outputs of individual magnetometers within the network,
constrain the parameter space describing the kinetic mixing of dark photons over the mass range from 4.1\,feV to 2.1\,peV,
substantially improving previous dark-photon dark matter limits.
Further optimization holds promise for accessing an unexplored parameter domain beyond the astrophysical limitations imposed by anomalous plasma heating and cosmic microwave background distortion\,\cite{caputo2021dark}.

\section{Summary and perspectives}
\label{SecVI}



The accelerated progress in the domain of spin-based quantum sensors has heralded unprecedented opportunities for delving into the realm of exotic spin-dependent interactions beyond the SM.
With the ongoing evolution of spin-based quantum technologies, it is anticipated that spin sensors will increasingly serve as complementary tools to large-scale particle accelerators and direct particle detection approaches.
Hereinafter, we outline the challenges and prospects inherent in future explorations of these exotic interactions.
This encompasses the development of innovative principles for spin sensing, advancements in sensor techniques, and the evolution of theories.

Although there has been considerable progress, the full potential of spin sensors in detecting exotic spin-dependent interactions has yet to be fully realized.
Spin sensors exhibit an impressive standard quantum limit on measurement sensitivity, but there remains a significant gap between the current achievements and this theoretical threshold\,\cite{mitchell2020colloquium}.
For example, the best sensitivity achieved by the SERF magnetometer is currently at 0.16\,fT/Hz$^{1/2}$\,\cite{dang2010ultrahigh},
indicating a substantial opportunity for improvement when compared to the standard quantum limit, which is at the aT/Hz$^{1/2}$ level.
It is also essential to recognize that these standard quantum limits, as detailed in Sec.\,\ref{Sec:sensor}, represent a critical bottleneck in the pursuit of certain exotic spin-dependent interactions.
For example, in the case of QCD axions,
coupling constants typically scale inversely with the axion mass $m_a$.
At a mass of $m_a=10^{-12}$\,eV,
the coupling constant can be $g_{\mathrm{a N}}\leq 10^{-20}$,
corresponding to a magnetic field of 10$^{-10}$\,fT.
Reaching this level of sensitivity poses a significant challenge within the constraints of the standard quantum limit.
Therefore, to achieve enhanced fundamental sensitivity, the exploration of novel spin techniques or systems is imperative.

{Addressing this gap necessitates the innovation of new techniques to enhance signal response and conduct an exhaustive investigation into the sources of noise.
Firstly, to enhance the signal response, one effective method is to enhance the number of polarized spins.
This can be achieved by increasing the number of spins via high pressure gases or liquid-state or solid-state spin samples.
The polarization can be increased using optical pumping\,\cite{walker1997spin}, para-hydrogen induced polarization\,\cite{theis2011parahydrogen} and other hyperpolarization methods.
Another method is to extend the spin coherence time by coating techniques and quantum control methods, such as the DD coupling sequences.
For instance, these improvement can substantially elevate the amplification factor of magnetic or pseudomagnetic fields of the spin-based amplifiers\,\cite{jiang2021search,su2021search,wang2022limits,wang2023search}.
Secondly, there is a pressing need for suppressing various types of noises in the spin sensors.
The classical noise, such as magnetic field noise interference in experiments probing new physics, is of paramount importance. 
The exploration of new physical methodologies to naturally reduce sensitivity to magnetic noise is necessary.
Preliminary studies have highlighted the low-frequency magnetic noise insensitivity of SERF comagnetometers\,\cite{kornack2005nuclear}, marking an important milestone. 
More recent research into Fano effects has demonstrated the capability for self-compensation of magnetic noise across a broader frequency range\,\cite{jiang2023enhanced}.
As the classical noises are suppressed, the quantum noises such as photon-shot noise and spin-projection noise becomes dominant.
In order to surpass the standard quantum limit,
leveraging spin correlation or squeezing presents a promising path\,\cite{bao2020spin}.
However, achieving this goal necessitates significant advancements in the generation of entangled or squeezed states among large particle assemblies. 
Enhancements in the degree of spin squeezing and the extension of their coherence time are critical areas requiring breakthroughs.
Apart from improve current systems, another important way is to study new atomic, molecular, and condensed-matter systems that feature enhanced sensitivity beyond the current techniques, such as the Magnetic Needle Magnetometers\,\cite{band2018dynamics,kimball2016magnetic}.
These systems are proposed with a projected sensitivity far beyond current systems and their experimental demonstrations are ongoing.
These emerging avenues of research bear promising implications for future explorations in new physics experiments, potentially unlocking new realms of understanding in the field.}

The development of innovative detection methodologies is imperative to transcend the constraints imposed by current experimental configurations\,\cite{belenchia2022quantum,lu2022micius,el2020aedge}. Space-based quantum sensors are at the forefront of opening new vistas for the investigation of ultralight dark matter and the exploration of exotic spin-dependent forces.
A notable example is a recent proposal that highlights the potential of atomic clocks
onboard to the inner reaches of the
solar system\,\cite{tsai2023direct}.
This approach aims not only to detect a dark matter halo gravitationally bound to the Sun but also to observe spatial variations in the fundamental constants, potentially induced by alterations in the gravitational potential.
The sensitivity that space-based clocks could achieve in detecting a Sun-bound DM halo vastly surpasses that of terrestrial clocks, expanding the scope of detection by several orders of magnitude.
While this proposal primarily focuses on the utility of atomic clocks as quantum sensors, spin sensors emerge as equally promising for the investigation of axion-spin interactions within the Sun-bound DM halo.
Moreover, the deployment of quantum sensors on space stations is poised to significantly enhance sensitivity to exotic spin-dependent forces, especially those dependent on velocity,
enhancing detection capabilities by several orders of magnitude\,\cite{Huangsearch2024}.
As space technology continues to evolve, space technology of spin sensors are gradually maturing, thereby creating new opportunities for the future detection of new physics.

Building on the advances and challenges outlined in the exploration of exotic spin-dependent interactions beyond the SM, the theoretical exploration in particle physics, particularly concerning axions and dark photons, holds immense promise. A pivotal objective lies in comprehensively exploring the parameter space of the QCD axion, where the product of the axion mass $m_a$ and the decay constant $f_a$ is approximately $(0.1\,\text{GeV})^2$. This decay constant not only signifies the ultraviolet scale of Peccei-Quinn symmetry breaking\,\cite{Peccei:1977hh,Peccei:1977ur} but also paves the way for understanding the broader implications of axions in cosmology\,\cite{OHare:2024nmr}, particularly regarding the formation of axion dark matter. The predicted coupling of axions to fermions, $g_{a\psi} \sim m_\psi/f_a$ where $m_\psi$ denotes the fermion mass, encompasses a dimensionless coefficient associated with the ultraviolet model. 

The theoretical framework involves various models that may facilitate diverse interactions between SM fermions and ultralight bosons, with boson masses spanning several orders of magnitude. Understanding these ultraviolet origins is crucial, particularly for enhancing the motivation and direction of experimental searches. Therefore, it is essential to continuously advance theoretical investigations on exotic spin-dependent interactions, specifically pinpointing which interactions should be prioritized for groundbreaking efforts. Strengthening theoretical research further enables the provision of more precise theoretical expectations to guide the design of experiments and the analysis of data, thereby fostering progress in the field of new spin interactions.  Meanwhile, employing spin sensors in research helps explore the extensive range of potential interactions between new bosonic particles and SM fermions, thus broadening the scope of searches for new particles and forces beyond the SM.

Spin sensors can address another pivotal question---the nature of dark matter, particularly if it consists of new bosonic fields. Depending on the production mechanism and the dynamics within galaxies, the local dark matter profile may manifest either as an isotropic, coherently oscillating background or as a directional stream wave\,\cite{OHare:2017yze,Foster:2017hbq,Knirck:2018knd}. Additionally, dark matter may condense into local clumps or form topological defects. The detectable signals from these phenomena are typically proportional to the boson's wavefunction and, in the case of axion-fermion interactions, to the spatial gradient. Thus, spin sensors are crucial for directly probing the characteristics of such background fields, thereby enhancing our understanding of both the microscopic nature and macroscopic properties of dark matter. Furthermore, by initiating searches for various bosonic sources, spin sensors can open a new avenue in multimessenger astronomy and cosmology.



We thank Dong Sheng, Teng Wu, S.-B. Zhang, Qing Lin, and Tian Xia for valuable discussions.
This work was supported by the Innovation Program for Quantum Science and Technology (Grant No.\,2021ZD0303205),
National Natural Science Foundation of China (Grants Nos.\,T2388102, 11927811, 12150014, 12274395, 12261160569), Youth Innovation Promotion Association (Grant No.\,2023474), and Chinese Academy of Sciences Magnetic Resonance Technology Alliance Research Instrument and Equipment Development/Functional Development (Grant No. 2022GZL003).
Y.C. is supported by VILLUM FONDEN (Grant No. 37766), by the Danish Research Foundation, and under the European Union’s H2020 ERC Advanced Grant “Black holes: gravitational engines of discovery” grant agreement no. Gravitas–101052587, by the Munich Institute for Astro-, Particle and BioPhysics (MIAPbP) which is funded by the Deutsche Forschungsgemeinschaft (DFG, German Research Foundation) under Germany´s Excellence Strategy – EXC-2094 – 390783311, by the European Consortium for Astroparticle Theory in the form of an Exchange Travel Grant, and by FCT (Fundação para a Ciência e Tecnologia I.P, Portugal) under project No. 2022.01324.PTDC.





\bibliographystyle{naturemag}
\bibliography{mainrefs1}

\end{document}